\documentclass[twocolumn]{aastex63}
\usepackage{graphicx}
\usepackage[normalem]{ulem}
\usepackage[nointegrals]{ wasysym }
\usepackage{amsmath}

\def\lp{\left(}
\def\rp{\right)}
\def\lb{\left[}
\def\rb{\right]}
\def\t{\text}

\received{date}
\revised{date}
\accepted{\today}

\shorttitle{Protostellar CRs}
\shortauthors{Fitz Axen et al.}

\begin{document}

\title{Transport of Protostellar Cosmic Rays in Turbulent Dense Cores}

\correspondingauthor{Margot Fitz Axen}
\email{fitza003@utexas.edu}

\author[0000-0001-7220-5193]{Margot Fitz Axen}
\affiliation{Department of Astronomy, The University of Texas at Austin, 2515 Speedway, Stop C1400, Austin, Texas 78712-1205, USA.}

\author[0000-0003-1252-9916]{Stella S. S. Offner}
\affiliation{Department of Astronomy, The University of Texas at Austin, 2515 Speedway, Stop C1400, Austin, Texas 78712-1205, USA.}
\affiliation{Center of Planetary Systems Habitability, The University of Texas at Austin, Austin TX 78712, USA}

\author[0000-0003-4224-6829]{Brandt A. L. Gaches}
\affiliation{I. Physikalisches Institut, Universit\"{a}t zu K\"{o}ln, Z\"{u}lpicher Stra{\ss}e 77, 50937, K\"{o}ln, Germany}
\affiliation{Center of Planetary Systems Habitability, The University of Texas at Austin, Austin TX 78712, USA}

\author[0000-0003-2624-0056]{Chris L. Fryer}
\affiliation{Center for Theoretical Astrophysics, Los Alamos National Laboratory, Los Alamos, NM 87544, USA}

\author{Aimee Hungerford}
\affiliation{Center for Theoretical Astrophysics, Los Alamos National Laboratory, Los Alamos, NM 87544, USA}

\author[0000-0003-1572-0505]{Kedron Silsbee}
\affiliation{Max-Planck-Institut f\"{u}r Extraterrestrische Physik, Giessenbachstra{\ss}e, 85748, Garching, Germany}

\begin{abstract}
    Recent studies have suggested that low-energy cosmic rays (CRs) may be accelerated inside molecular clouds by the shocks associated with star formation. We use a Monte Carlo transport code to model the propagation of CRs accelerated by protostellar accretion shocks through  protostellar cores. We calculate the CR attenuation and energy losses and compute the resulting flux and ionization rate as a function of both radial distance from the protostar and angular position. We show that protostellar cores have non-uniform CR fluxes that produce a broad range of CR ionization rates, with the maximum value being up to two orders of magnitude higher then the radial average at a given distance. In particular, the CR flux is focused in the direction of the outflow cavity, creating a 'flashlight' effect and allowing CRs to leak out of the core. The radially averaged ionization rates are less than the measured value for the Milky Way of $\zeta \approx 10^{-16} \rm s^{-1}$; however, within $r \approx 0.03$ pc from the protostar, the maximum ionization rates exceed this value. We show that variation in the protostellar parameters, particularly in the accretion rate, may produce ionization rates that are a couple of orders of magnitude higher or lower than our fiducial values. Finally, we use a statistical method to model unresolved sub-grid magnetic turbulence in the core. We show that turbulence modifies the CR spectrum and increases the uniformity of the CR distribution but does not significantly affect the resulting ionization rates.
\end{abstract}


\section{Introduction}
\label{section:intro}

Cosmic rays (CRs) are the primary drivers of ionization inside the densest regions of molecular clouds where UV radiation cannot penetrate \citep{dalgarno_2006_crs, padovani_2020_review}. The degree of ionization in star-forming regions is critical to setting the atomic and molecular abundances  \citep{wakelam_2012_astrochemistry}. Ionization also impacts the gas dynamics inside the cloud, as ionized species couple to the magnetic field and resist collapse while neutral molecules decouple from the field through ambipolar diffusion \citep{fiedler_1993_ambipolar}. Ionization is especially crucial to the evolution of circumstellar disks, as the level of ionization determines whether the  Magnetorotational Instability (MRI) is active \citep{balbus_1991_mri, gammie_1996_mri}. The MRI impacts the dynamics of angular momentum transport, accretion, and the ability of the disk to form planetesimals \citep{dutrey_2014_protoplanetary}.

 The level of cosmic ray ionization is usually expressed quantitatively as the Cosmic Ray Ionization Rate (CRIR), which is the rate of CR ionizations per hydrogen molecule. The mean galactic ionization rate is $\zeta \approx 10^{-16} \text{s}^{-1}$ \citep{indriolo_2012_crir}; however, observations along different sightlines suggest higher rates towards the Galactic center \citep{indriolo_2015_herschel}. CR ionizations produce molecules such as $\t{OH}^+$, $\t{H}_2\t{O}^+$, and $\t{H}_3\t{O}^+$ in diffuse interstellar clouds, so observations of these molecules indicate the abundance of CRs and can be used to derive the CRIR \citep{indriolo_2015_herschel}.

While high energy CRs (E $>$ 100 GeV) are primarily accelerated to relativistic energies by supernovae and active galactic nuclei \citep{morlino_2017_supernovae, berezhko_2008_AGNs}, recent studies suggest that the shocks associated with star formation also have the right conditions to accelerate CRs with lower energies of E $<$ 100 GeV. For example, observations of $\t{HCO}^+$ and $\t{N}_2\t{H}^+$ along sightlines towards the protostars OMC-2 FIR 4 and L1157-B1 indicate values higher than the mean CRIR \citep{ceccarelli_2014_omc, fontani_2017_omc2, podio_2014_shock}. \cite{favre_2018_omc} measured a value of $\zeta \approx 10^{-14} \t{s}^{-1}$ towards OMC-2 FIR 4, two orders of magnitude higher than the Milky Way value. In general, differences in the ionization rate in certain regions may be explained by the magnetic field morphology in those regions. Measurements of the magnetic fields in young protostellar systems have found that they trace an hour-glass-like shape \citep{girart_2006_bfield}, though this is not necessarily ubiquitous. To leading order, this field geometry does not significantly impact the CR flux \citep{fatuzzo_2014_clouds}; however, the environments surrounding young stellar objects are expected to be turbulent. Theoretical studies have found that in the presence of `magnetic pockets' - regions with a locally reduced magnetic field strength that naturally occur in magnetized turbulence - the CRIR may be significantly lower \citep{fatuzzo_2014_clouds, silsbee_2018_turbulence}. Consequently the elevated observed ionization rates suggest a local mechanism is acting to accelerate CRs inside molecular clouds.

Another signature of CR acceleration is non-thermal synchrotron emission, which is caused by relativistic electrons gyrating around a magnetic field. Evidence of non-thermal emission has been linked to shocks in protostellar jets \citep{anglada_2018_jets}. Synchrotron emission has been observed from the HH 80-81 jet and a jet in the star forming region G035.02+0.35, which are both associated with young massive stars \citep{kamenetzky_2017_hh80, sanna_2019_protostellar}. Synchrotron emission has also been detected from the jet of the lower mass young sunlike star DG Tau \citep{ainsworth_2014_dgtau}. These results further support the evidence that the formation of young stars generates low-energy CRs. 

Theoretical studies have modeled the acceleration of CRs associated with protostellar shocks \citep{padovani_2015_protostars, padovani_2016_protostars}.  \cite{padovani_2016_protostars} developed a model of protostellar CR proton and electron acceleration at three potential sites; accretion flow shocks in collapsing envelopes, jet shocks, and surface accretion shocks.  \cite{gaches_2018_exploration} extended the surface accretion shock predictions to include more detailed models of the shocks, focusing on connecting the properties of the shock and the resulting CR spectrum to the properties of the forming star. These authors also investigated the attenuation of the CR spectrum in clouds, dense cores and disks \citep{padovani_2016_protostars, padovani_2018_disks, gaches_2018_exploration, offner_2019_disks}. However, they used analytic models that made simplifications assuming isotropic diffusion for the protostellar core properties.

Protostellar cores exhibit a number of asymmetries that likely influence CR propagation. Both the magnetic field and the density of the gas influence CR propagation; the magnetic field determines the direction of their travel, and the gas density determines how much energy they lose. Cores are often prolate and elongated along the direction of filaments \citep{myers_1991_cores, nutter_2007_orion}. Cores, particularly those in higher mass star forming regions like Orion are also turbulent \citep{kirk_2017_orion}.
In the star formation process high momentum outflowing material coupled to magnetic field lines sweeps up and removes the dense gas in the core, creating an outflow cavity \citep{bally_2016_outflows}. Significant deviations in the symmetry of the gas density and magnetic field around the star will produce different CR flux spectra and ionization rates in different regions. Therefore, having a fully consistent treatment of the geometry of the gas density and magnetic field is crucial to accurately modeling CR propagation within protostellar disks, cores, and molecular clouds.

In this work we explore the propagation of CR protons accelerated by a protostellar accretion shock from the star's surface to the edge of the protostellar core. We model the propagation of the CRs numerically, using previous simulations of forming protostars as inputs for the magnetic field and gas density in the core. We use these results to compute the attenuation of the CR energy spectrum and ionization rate as a function of radius from the protostar and solid angle, going beyond the symmetric limitations of past theoretical studies. We explore the impacts of different protostellar ages and amount of magnetic turbulence in the core on the propagation of the CRs. In Section \ref{section:methods} we describe the model we use and our numerical methods. In Section \ref{section:results} we discuss the results of the CR propagation and ionization rates in the core. Finally, in Section \ref{section:discussion} we compare our results to observations and discuss the further implications of our work.

\section{Methods}
\label{section:methods}

\subsection{Model for CR Acceleration in Protostellar Accretion Shocks}
\label{subsection:model}

We consider CR acceleration through Diffusive Shock Acceleration (DSA), or first order Fermi acceleration, in which a charged particle encounters a shock and scatters back and forth across the shock width due to magnetic fluctuations, gaining energy as it does so \citep{drury_1983_dsa}. The shock must have the right conditions for the particle to be accelerated efficiently without losing too much energy or leaving the shock. Specifically, the shock must be super-sonic ($U_s>c_s$) and super-Alfv\'enic ($U_s>v_{\rm A}$), where $U_s$, $c_s$, and $v_{\rm A}$ are the shock velocity, sound speed, and Alfv\'en speed, respectively. The sound speed in the shock is given by \citep{padovani_2019_synchrotron}
\begin{equation}
    c_s=9.1\lb\gamma_{\t{ad}}(1+x)\lp\frac{T_s}{10^4 \t{K}}\rp\rb^{0.5} \t{km s}^{-1}, 
\end{equation}
where $\gamma_{\t{ad}}=5/3$ is the adiabatic index, $T_s$ is the pre-shock temperature, and $x$ is the ionization fraction ($x \approx 1$ at the surface of the star). The Alfv\'en velocity is given by 
\begin{align}
v_{\rm A} &= \frac{B_s}{\sqrt{4\pi\rho_s}} \nonumber \\
       &=2.2\times10^{-2}\lp \frac{n_s}{10^6 \t{cm}^{-3}}\rp^{-0.5}\lp\frac{B_s}{10 \mu \t{G}}\rp \t{km s}^{-1},
\end{align}
where $B_s$ is the shock magnetic field, $\rho_s$ is the shock mass density, and  $n_s=\frac{\rho_s}{\mu_{\t{I}}m_{\t{H}}}$ is the shock number density. Here  $\mu_{\t{I}}$ is the mean molecular weight, and $m_{\t{H}}$ is the atomic hydrogen mass. We assume a fully ionized gas and use a value of $\mu_{\t{I}}=0.6$. 

To compute the initial CR spectrum at the protostellar surface we follow the analytical calculation done by \cite{gaches_2018_exploration},
which builds on the model from \cite{padovani_2016_protostars}. The properties of the spectrum are determined by a model for the protostellar accretion shock, which accelerates particles to relativistic energies. \cite{gaches_2018_exploration} developed a semi-analytic model for protostellar accretion and the corresponding CR spectrum as a function of protostellar mass $M_*$, protostellar radius $R_*$, and accretion rate $\dot{m}$, which we describe further in Section \ref{subsubsection:spectrum}. Their protostellar accretion shock model is parameterized by the shock velocity, temperature of the gas, and density of the gas \citep{hartmann_2016_accretion}. 

We assume the shock velocity, $U_s$, is approximately the free-fall velocity at the protostellar surface $R_*$:
\begin{align}
    U_s &= \sqrt{\frac{2GM_*}{R_*}} \nonumber \\
    &= 309\lp\frac{M_*}{0.5M_{\astrosun}}\rp^{0.5}\lp\frac{R_*}{2R_{\astrosun}}\rp^{-0.5} \t{km s}^{-1}.
\end{align}
In the strong shock regime, the post-shock temperature is given by \citep{hartmann_2016_accretion}:
\begin{align}
    T_s &= \frac{3}{16}\frac{\mu_{\t{I}}m_{\t{H}}}{k}U_s^2 \nonumber \\
    &= 1.302\times10^6\lp\frac{\mu_{\t{I}}}{0.6}\rp\lp\frac{U_s}{309 \t{km s}^{-1}}\rp^2 \t{K.}
\end{align}
 The shock density is given by:
\begin{align}
    \rho_s &=\frac{\dot{m}}{AU_s} \nonumber \\
    &= 8.387 \times 10^{-10} \lp\frac{\dot{m}}{10^{-5} M_{\astrosun} \t{yr}^{-1}}\rp \lp\frac{f}{0.1}\rp^{-1}
    \nonumber \\
    & \times \lp\frac{R_*}{2R_{\astrosun}}\rp^{-2}\lp\frac{U_s}{309 \t{km s}^{-1}}\rp^{-1} \t{g cm}^{-3},
\end{align}
where $A=4\pi fR_*^2$ is the area of the accretion columns and $f$ is the filling fraction of the accretion columns on the surface of the protostar.

The energy spectrum of cosmic rays produced by DSA forms a power-law distribution in momentum space:
\begin{equation}
    f(p)=f_0\lp\frac{p}{p_{\t{inj}}}\rp^{-\alpha}, 
\label{equation:spectrum_power}
\end{equation}
where $f_0$ is a normalization factor, $p_{\t{inj}}$ is the injection momentum of thermal CR particles, and $\alpha$ is the power-law index, which is related to the shock compression ratio $r_s$ by $\alpha=\frac{3r_s}{r_s-1}$. The flux spectrum of the CRs is given by:
\begin{equation}
    j(E)=\frac{v(E)N(E)}{4\pi}, 
\end{equation}
where $v(E)=\beta c$ is the velocity as a function of energy, $E$, and the energy number density distribution is given by:
\begin{equation}
    N(E)=4\pi p^2 f(p) \frac{dp}{dE}. 
\end{equation}
The energy spectrum is defined over a range $E_{\t{inj}} < E < E_{\t{max}}$, where $E_{\t{inj}}$ is the injection energy of thermal CRs and $E_{\t{max}}$ is the maximum energy for which CRs can be efficiently accelerated by DSA. The injection energy, $E_{\t{inj}}$, depends on the strength of the shock and the hydrodynamic properties, while the maximum energy $E_{\t{max}}$ is determined by the magnetic field, ionization fraction, and shock acceleration efficiency \citep{gaches_2018_exploration}. The details describing the calculation of these energy parameters are outlined in \cite{gaches_2018_exploration} and \cite{padovani_2016_protostars}.

\subsection{CR Propagation}
\label {subsection:propagation}

\subsubsection{Numerical Method}
\label{subsubsection:numerical}

Our aim is to model the attenuation of the CR spectrum in the protostellar core after the CRs are accelerated at the surface by the accretion shock. Past studies have calculated the attenuation of the spectrum analytically by making assumptions about the properties of the core. Here, we instead follow the propagation of each individual CR particle through a realistic magnetic field and density configuration. This method has the advantage that it accounts for spatial variations in the core properties and allows us to calculate attenuated spectra and ionization rates in different regions of the core. 

To compute the CR propagation we use a three-dimensional, time-independent Monte Carlo radiation transport code designed for the propagation of CR protons \citep{fitzaxen_2021_CRs}. A CR source can be placed at any location in the grid and emits particles in random initial directions. The code propagates particles through a three dimensional grid of cubic cells with physical properties (density, bulk magnetic field) that are constant within each cell. The particles travel through the grid from their starting location until they exit one of the grid boundaries.

The code determines the propagation of particles by approximating the Lorentz equation of motion. The Lorentz equation describes the motion of particles spiraling around a magnetic field line, with a gyroradius of motion $r_g$. For a charged particle moving with velocity, $\textbf{v}(t)$, through a magnetic field, $\textbf{B}(\textbf{x})$, the Lorentz equation is:
\begin{equation}
    \label{eq:Lorentz}
    \frac{d\textbf{u}(t)}{dt} =\frac{\gamma\beta}{r_g}(\textbf{v}(t)\times \hat{\textbf{B}}(\textbf{x})),
\end{equation}
where $\textbf{u}(t)=\gamma \textbf{v}(t)/c$, $c$ is the speed of light, and $\gamma$ and $\beta$ are the Lorentz factors, which are related to the velocity of the particle by $\gamma=1/\sqrt{1-v^2/c^2}$ and $\beta=v/c$. The gyroradius for a particle with charge, $e$, is given by
\begin{equation}
    r_g= \frac{p c}{e B} = 2.2 \times 10^{-3} \beta \lp\frac{E_{\rm tot}}{\rm GeV}\rp\lp\frac{B}{100 \mu \rm G}\rp^{-1} \rm au,
\end{equation}
where $p=mv\gamma$ is the momentum of the particle, $m$ is the mass of the particle, and $E_{\rm tot}=E+m_c^2$. For a proton with a kinetic energy of 1 GeV in a magnetic field of strength 100 $\mu$G, typical of our problem, $r_g$ is much smaller than the grid cells we model.

 Solving Equation $\ref{eq:Lorentz}$ is computationally expensive because of the number of directional changes the particle makes. Therefore, the code makes the common approximation that the particle simply follows the magnetic field line that it experiences \citep{harding_2016_CRs, fitzaxen_2021_CRs}. This is a valid approximation as long as $r_g$ changes on a much smaller scale than the magnetic field because of the assumption that solving Equation \ref{eq:Lorentz} would be unlikely to move the particle into a region with a different magnetic field. Under this approximation, the particle does not change direction until the direction of the local magnetic field changes. For a particle traveling for a timestep $\Delta t$, the particle's position at time $t+\Delta t$ can be calculated as: 
\begin{equation} \label{eq:transpos}
\textbf{x}(t+\Delta t)= \textbf{x}(t)+ \frac{(\textbf{v}(t) \cdot \textbf{B})}{B}\hat{B} \Delta t,
\end{equation}
where $\textbf{x}$ and $\textbf{v}$ are the particle's position and velocity at time $t$. The particle travels either parallel or antiparallel to the direction of the magnetic field line.

The distance the particle travels in a single step $\Delta x=v \Delta t$ is sampled using the scattering path length $\lambda_{\rm sc}$, which is the expected distance the particle travels before the local magnetic field changes. For the scattering length we adopt the expression from Equation 3 in  \cite{fryer_2007_probing}:
\begin{equation}
    \label{eq:lambda_sc}
    \lambda_{\t{sc}}=c(\gamma\beta)^{2-q}f(\beta_A)\frac{8}{\pi(q-1)\sigma^2ck_{\t{min}}}\lp\frac{ ck_{\t{min}}}{\Omega}\rp^{2-q}. 
\end{equation}
Here the quantity $f(\beta_A)$ is a function of the Alfv\'en velocity $\beta_A$ in the plasma, $q$ is the index of the power spectrum for the turbulence, and  $\sigma^2=(\delta B/B_0)^2$ is the power of the turbulent fluctuation $\delta \textbf{B}$ relative to the bulk magnetic field $\textbf{B}_0$. The bulk magnetic field $\textbf{B}_0$ is the mean magnetic field over some distance, in our case, a grid cell. The minimum wavenumber of the magnetic field $k_{\t{min}}$ is related to the maximum scale length of the magnetic field by $k_{\t{min}}=2\pi/\lambda_{\t{max}}$, where $\lambda_{\t{max}} \approx L_c$, the correlation length of the magnetic field.

Particles with a scattering length $\lambda_{\t{sc}}$ follow an exponential probability distribution for their motion, $P(\Delta x)=e^{-\Delta x/\lambda_{\t{sc}}}$. Solving this equation for the cumulative distribution function (CDF) gives $\chi = 1-e^{-\Delta x/\lambda_{\t{sc}}}$, where $\chi$ is a random value sampled between 0 and 1. This can be rearranged to sample a distance traveled for that step as:
\begin{equation} \label{eq:transdis} 
\Delta x= -\text{ln}(\chi)\lambda_{\t{sc}}.
\end{equation}
If $\delta \textbf{B}=0$, there is no magnetic turbulence, and $\lambda_{\t{sc}}$ is infinite. In this case, $\Delta x$ is limited by the size of the grid cell as the particle will change direction only when it encounters a new value of $\textbf{B}_0$. 

For each particle step through the grid, (i.e, each recalculation of $\Delta x$) the code samples the magnetic field and computes the path length for the next step. The method checks whether traveling this distance would take the particle into a different grid cell where there is a different bulk magnetic field. If it does, the particle travels only to the edge of the cell, even if this distance is less than $\Delta x$. When it enters a new cell its path length and travel distance are resampled and its motion continues along the previously determined direction \citep{harding_2016_CRs, fitzaxen_2021_CRs}.

The code also calculates the energy losses the CR particles experience after every step. As CRs propagate through the gas, they undergo interactions that cause them to lose energy and change the shape of the flux spectrum. These energy losses can be represented as a loss function, $L(E)$, which is defined as: 
\begin{equation}
    \label{eq:loss}
    L(E)=\frac{1}{n(H_2)}\frac{dE}{dx}=\frac{1}{n(H_2) v}\frac{dE}{dt},
\end{equation}
where $n(H_2)=\rho/2.8m_{\t{H}}$ is the local number density of molecular hydrogen.

\subsubsection{Magnetic Field Treatment}
\label{subsubsection:bfield}

The coupling between CRs and the mean magnetic field is determined by the amount of magnetic turbulence \citep{rodgers-lee_2017_CRs, thomas_2019_hydrodynamics, bai_2019_magnetohydrodynamic}. Dense protostellar cores are expected to be at least sub-sonically turbulent, while outflows and disk winds can drive additional turbulence \citep{bai_2016_disks, offner_2017_impact}. Additionally, MHD turbulence can be driven by streaming instabilities caused by anisotropies in the CR distribution function \citep{morlino_2015_instability, padovani_2020_review}. The amount of turbulence is relatively unknown, however, and depends on parameters such as the core formation process, infall from the envelope, and gas properties such as magnetic field strength and ionization.

Particles in a turbulent magnetic field approximate a random walk. This means that for many particles, the positions will be isotropically distributed in angle around their source at distances much further away than a scattering length. For this reason, many commonly used CR propagation codes such as GALPROP assume that CRs propagate entirely due to spatial and momentum diffusion \citep{strong_1998_galprop}. In the absence of energy losses, the attenuation of the CR flux in a turbulent magnetic field can be solved for analytically: $j(E,r)\propto j_{\rm in}(E) r^{-1}$, where $j_{\rm in}(E)$ is the flux at the stellar surface $R_*$. The opposite extreme is the free streaming solution, in which particles are in a completely isotropic magnetic field and do not change direction at all. In the free streaming limit, without energy losses, the flux attenuates as $j(E,r)\propto j_{\rm in}(E) r^{-2}$. 

In \cite{fraschetti_2018_mottled}, the authors perform test-particle numerical simulations of protons propagating in a T-Tauri stellar wind. They include a two component treatment for their magnetic field structure- a large scale component, $\boldsymbol{B_0}(x)$, that is generated by MHD simulations and is constant within each grid cell, and a small-scale turbulent component, $\boldsymbol{\delta B}(x)$, which changes on scales smaller than a grid cell. Following \cite{giacolone_1999_turbulence}, they explicitly calculate the direction of the turbulent component for every particle step as a sum over plane waves with randomly sampled polarizations and phases using a Kolmogorov power spectrum. They then use this to calculate the total magnitude and direction of the magnetic field as $\boldsymbol{B}(x) = \boldsymbol{B_0}(x)+ \boldsymbol{\delta B}(x)$, and follow particle propagation along this field line.

While this method for modeling turbulence is widely used, recomputing the turbulent field at every particle step is computationally expensive if the path length is much smaller than the grid. We follow \cite{fraschetti_2018_mottled} by modeling the magnetic field as having both a bulk component $\boldsymbol{B_0}(x)$ and a turbulent component $\boldsymbol{\delta B}(x)$. Unlike \cite{fraschetti_2018_mottled}, however, we treat the turbulent component $\boldsymbol{\delta B}(x)$ statistically, which allows us to examine a range of behaviors between strong turbulence and no turbulence. By choosing to use an alternative method for CR transport, we ensure that we can run many particles through the full extent of our grid. 

To determine the particle travel direction, we use a numerical method in which the particles either follow the bulk magnetic field direction or approximate random walk motion \citep{fitzaxen_2021_CRs}. For every step, we sample a random number $\chi$ between 0 and 1 and compare that to the regular field amplitude ratio $b=B_0 (x) / B(x) $, where $B(x)=B_0(x)+\delta B(x)$ is the sum of the magnitudes of the two components and $b$ is related to the turbulent power by $\sigma^2=(\frac{1}{b}-1)^2$. If $\chi>b$, then the particle travel direction is sampled randomly to approximate random walk motion. If $\chi<b$, then the particle follows the bulk magnetic field with one of two directions (either parallel or antiparallel to it) based on the velocity of its previous step. For example, if the amplitudes of the global and turbulent field are the same ($\sigma^2=1$), then $b=0.5$, and a particle will have a 50 \% probability of following the direction of $B_0$ and a 50 \% probability of sampling a random direction.

We do not have a simple method for testing our two component magnetic field treatment. However, we did test the behavior of our code in the free streaming and diffusion limits. First, we tested the behavior of the code in these limits without energy losses and verified that we reproduced the analytical solutions for the flux attenuation. We also tested both of these limits with energy losses and compared the results to a model developed by \cite{silsbee_2019_model}. These tests are described further in Appendix \ref{appendix_subsection:tests_losses}.

\subsection{Initial Conditions}
\label{subsection:initial}

\subsubsection{Gas Distribution}
\label{subsubsection:gas}

We take the magnetic field and density grid inputs and the initial conditions for the protostar from the magneto-hydrodynamic simulations performed by \cite{offner_2017_impact}. 
This study modeled the collapse of a dense core and the subsequent formation of a protostar for different initial magnetic field strengths and core masses. The simulations use the {\sc orion2} adaptive mesh refinement code \citep{li_2012_orion}. {\sc orion2} is a code which includes ideal magnetohydro-dynamics, self-gravity, gray radiative transfer, radiative feedback, and star particles with a model for protostellar evolution. Protostars form in the simulations when the density in a cell becomes high enough to violate the Jeans condition \citep{krumholz_2004_sink}. The protostar interacts with its surroundings by accreting, radiating, and launching gas outflows \citep{offner_2009_radiative, krumholz_2004_sink, cunningham_2011_outflows}. The accretion shock onto the protostar is our CR source as described in \cite{gaches_2018_exploration}.

The domain extent of the MHD simulations is $L=0.26$ pc, twice the initial diameter of the core, and the basegrid has $256^3$ cells. Five AMR levels are used to reach a maximum resolution of $\Delta x_{\rm AMR} \approx 26$ au. We use the density and magnetic field outputs at two different times to compute the CR propagation. The MHD code also includes a tracer fluid which tracks the density of the gas launched in the outflow. Following \cite{offner_2017_impact}, we define the tracer fraction $F_t=\rho_t/\rho$, which is the ratio of outflow gas density to total gas density in a grid cell. This ratio indicates the amount of material launched in the outflow and reflects the amount of mixing and entrainment of core material with the outflow. While most grid cells have values of $F_t < 0.01$, we define the outflow as those cells with $F_t > 0.1$. We examine cutoff values of $F_t=0.01$, $F_t=0.05$, and $F_{t,\rm out}=0.1$ to study regions of progressively wider collimation around the outflow cavity. While this quantity is not relevant to CR propagation, we use it to investigate how the outflow geometry influences the CR distribution.

\begin{figure*}[th!]
\centering
\includegraphics[width=0.98\linewidth]{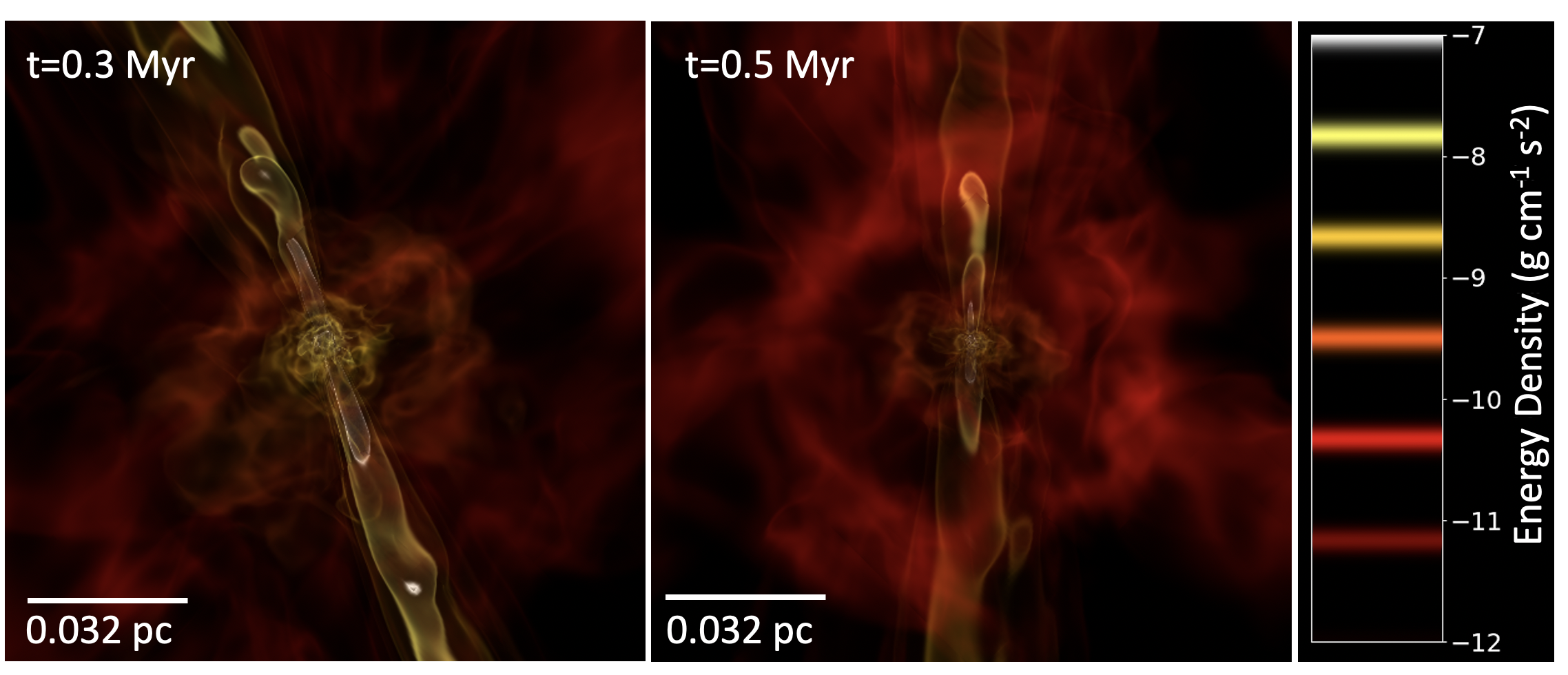}
\caption{Volume renderings of the gas energy density at $t=0.3$ Myr and $t=0.5$ Myr.}
\label{fig:gas_density}
\end{figure*}

\begin{figure*}[th!]
\centering
\includegraphics[width=0.98\linewidth]{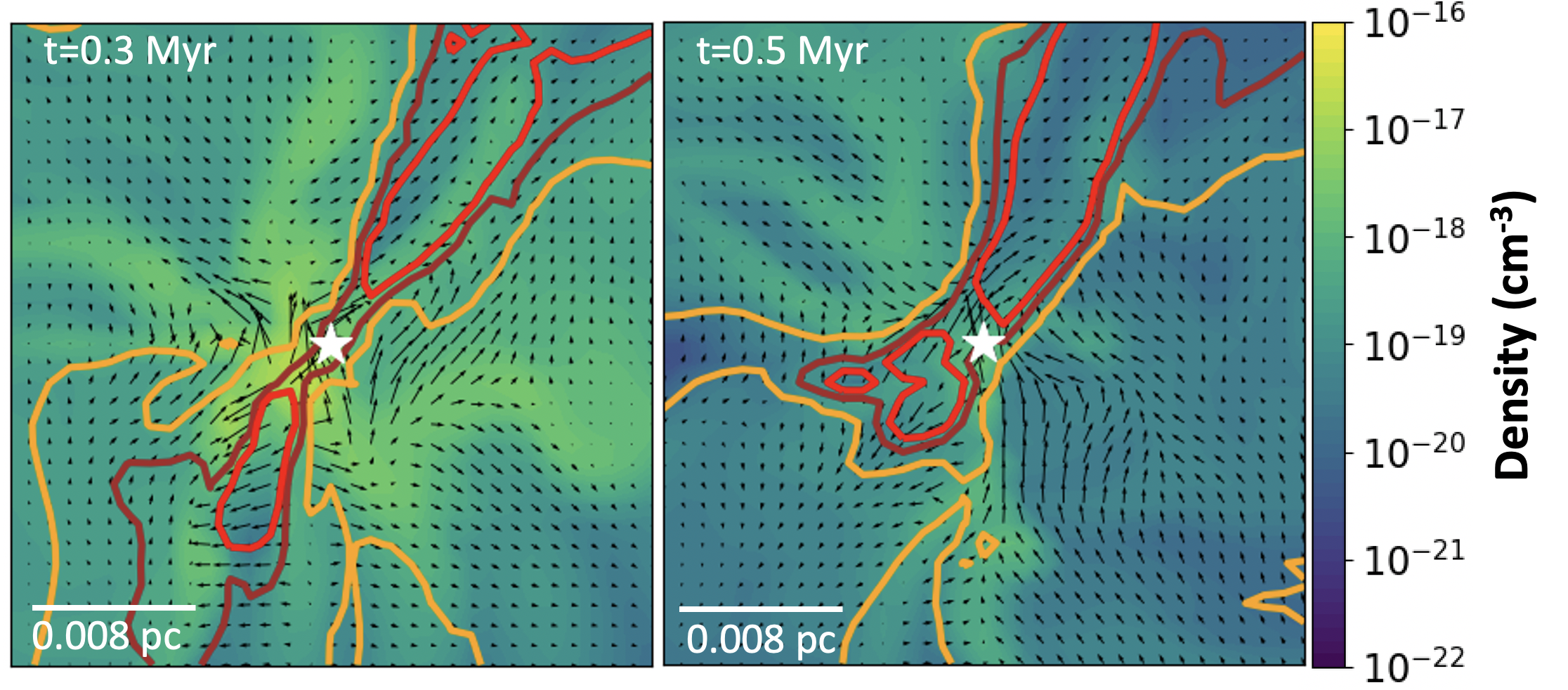}
\caption{A slice plot through the center of the star at $t=0.3$ Myr and $t=0.5$ Myr. Colorscale indicates the gas density. Contours outline the location of the outflow, with $F_t > 0.01$ in yellow, $F_t > 0.05$ in brown, and $F_t > 0.1$ in red. Vectors show the magnitude and direction of the magnetic field, which varies between $\approx$ 10 $\mu \rm G$ and $\approx$ 1 mG in our grid. The white star indicates the location of the protostar, which is located slightly off center of the grid and is discussed in Section \ref{subsection:properties}.}
\label{fig:bfield}
\end{figure*}

We use the outputs of the fiducial run from \cite{offner_2017_impact} and summarize the physical parameters in Table \ref{table:offner_initial_conditions}. We use the simulation outputs at two different times, $t=0.3$ Myr and $t=0.5$ Myr, in order to study the effect of protostellar age and changes to the core properties on CR propagation. The masses of the protostar at the two times are 0.83 $M\astrosun$ and 1.07 $M\astrosun$, at $t=0.3$ Myr and $t=0.5$ Myr respectively. Figure \ref{fig:gas_density} shows volume renderings of the gas energy density at these two times. Figure \ref{fig:bfield} shows slice plots of the grid density and magnetic field through the center of the star at the two times. The shape of the outflow cavity and the magnetic field configuration shown in these plots are mostly dependant on the initial magnetic field used in the \cite{offner_2017_impact} simulations. Figure \ref{fig:bfield} demonstrates that in these grids the direction of the magnetic field is asymmetric and correlated with the direction of the outflow cavity. Decreasing the initial magnetic field strength in the MHD simulations by a factor of a few would cause more asymmetry in the gas density and magnetic field direction \citep{lee_2017_bfield}. 

We discuss how the protostellar accretion rate and radius impact the CR spectrum in more detail in Section \ref{subsubsection:spectrum}.

\begin{table*} 
\centering
\begin{tabular}{ |c c c c c c c c c c| } 
\hline
 Model & $M_{\rm core} (M_{\astrosun})$ & $l_{\rm max}$ & $\Delta x_{\rm AMR}$ (au) & $B_z (\mu \rm G)$ & $M_{\phi} (M_{\astrosun})$ & $\mu_{\phi}$ & $M_{\rm cr, eff} (M_{\astrosun})$ & $\mu_{\phi, \rm eff}$ & $v_i \rm (km  s^{-1})$\\
 \hline
MP45 & 4.0 & 5 & 26 & 20.6 & 0.80 & 5 & 1.27 & 3.15 & 0.52 \\
 \hline
\end{tabular}
\caption{Shows the inputs used in the \cite{offner_2017_impact} simulations that we use. Columns are the model name, gas mass, maximum AMR level, cell size on the maximum level, magnitude of the initial magnetic field in the $z$ direction, the magnetic mass, mass-to-flux ratio, effective magnetic critical mass, effective mass-to-flux ratio, and initial rms turbulent velocity. The molecular core has an initial radius of $R_c = 0.065 \rm pc$, a uniform density, an initial temperature $T_c = 10 \rm K$, and an initial virial parameter of $\alpha_{\rm vir}=2.0$. For a detailed description of these paramaters, see \cite{offner_2017_impact}.}
\label{table:offner_initial_conditions}
\end{table*}

\subsubsection{Input CR Spectrum Properties}
\label{subsubsection:spectrum}

For the initial CR energy spectrum at each time, we use the accretion shock model outlined in Section \ref{subsection:model} which was developed by \cite{gaches_2018_exploration}. In their model, the initial spectrum of accelerated CRs depends on the mass, radius, and accretion rate of the protostar, which evolve during the \cite{offner_2017_impact} simulations as the star forms. It also depends on the filling fraction $f$, stellar magnetic field $B_*$, and shock efficiency $\eta$.

The mass, radius, and accretion rate of the protostar are determined by a model for protostellar evolution. The radius is a somewhat uncertain parameter and may not be single valued with mass due to the potential influence of underlying factors on the evolution, such as initial conditions and the accretion details. Additionally, there is uncertainty in the protostellar evolution model, which depends on the accretion history, assumed initial radius and accretion efficiency \citep{hosokawa_2011_model}.The protostellar radius adopted by the star particles in {\sc orion2} depends on the accretion history and is calculated as described in \cite{offner_2009_radiative}. Given uncertainties in pre-main sequence models, protostellar radii are uncertain to factors of $\sim$ 2. 

The accretion rate is known to be variable, and often fluctuates over small timescales during the \cite{offner_2017_impact} simulations, varying up to an order of magnitude. This is consistent with the expectation that young stars undergo accretion bursts \citep{audard_2014_ppvi}. The area of the accretion columns is $A=4\pi fR_*^2$, where $f$ is the accretion flow filling fraction \citep{gaches_2018_exploration}. The value of $\dot{m}$ typically scales with $f$, as faster accreting protostars have larger accretion columns. Additionally, for a star with a fixed mass, the stellar radius is higher for higher internal entropy; therefore, we expect the value of $R_*$ to scale with $\dot{m}$ as well \citep{hosokawa_2011_model}.

 The three most uncertain parameters in the model are the filling fraction $f$, protostellar magnetic field $B_*$, and shock efficiency parameter $\eta$. Studies of T Tauri stars infer a magnetic field of a few G to a few kG \citep{krull_2007_measurements, donati_2011_bfield, johnstone_2014_bfield}. This is a later stage of evolution than we consider for our simulations, but these studies suggest reasonable values to assume for protostellar magnetic field strengths. The SPIRou spectrograph \citep{spirou_2014} aims to study stellar magnetism at earlier stages and should provide better values of $B_*$ for future studies. The shock efficiency parameter represents the fraction of particles accelerated by the shock, and is constrained by the temperatures associated with protostellar shocks to be $\approx 10^{-6}- 10^{-4}$ \citep{padovani_2016_protostars}. A filling fraction of $f=0.1$ represents a moderately accreting protostar, but this value may be higher for younger sources and lower for older sources \citep{hartmann_2016_accretion}. Finally, for the shock compression ratio, we use a value of $r_s \approx 2$, which is appropriate for the high temperatures near a protostar \citep{padovani_2016_protostars}. Further discussion of these parameters can be found in \cite{gaches_2018_exploration} and \cite{padovani_2016_protostars}.
 
For our fiducial runs, we use the values of the radius and accretion rate at $t=0.3$ Myr and $t=0.5$ Myr given by the output of the \cite{offner_2017_impact} simulations. We adopt values of $f=0.1$, $B_*=10$ G, and $\eta=10^{-5}$ following \cite{gaches_2018_exploration}. Then, for each time, we study the impact of varying $R_*$, $\dot{m}$, $f$, and $\eta$ within the limits discussed to produce both a high input CR spectrum and a low input CR spectrum. The accretion rate is the dominant parameter in each model, as varying the accretion rate has the largest impact on the spectrum; therefore, these variations can be thought of as 'high accretion' and 'low accretion' models. We do not vary the stellar magnetic field $B_*$ because it has a negligible impact on the magnitude of the spectrum. The parameter values for all of the runs are given in Table \ref{table:star_parameters}.

The values of the stellar parameters also impact the maximum energy of the CRs, $E_\t{max}$, in the initial spectrum. As shown in \cite{gaches_2018_exploration}, the maximum energy is determined by the outcomes of three different processes; upstream escape, wave damping and collisions; which of these effects dominates is highly parameter dependant. Here, we choose to set it at $E_{\t{max}}=3$ GeV for all of the initial spectra. This is roughly the median of the range of values that $E_{\t{max}}$ can take, about 1-10 GeV \citep{gaches_2018_exploration}. Moreover, the ionization by CRs with energies above 100 MeV is insignificant due to their low ionization cross section (see \cite{padovani_2009_cosmic}, Figure 1), so this choice does not significantly impact our results. Similarly, particles with energies below 100 keV also have a low cross section \citep{padovani_2009_cosmic} and lose energy too rapidly to contribute significantly to ionization; thus, we set $E_{\t{min}}= 100$ keV.  

\begin{figure}[th!]
\centering
\includegraphics[width=0.49\textwidth]{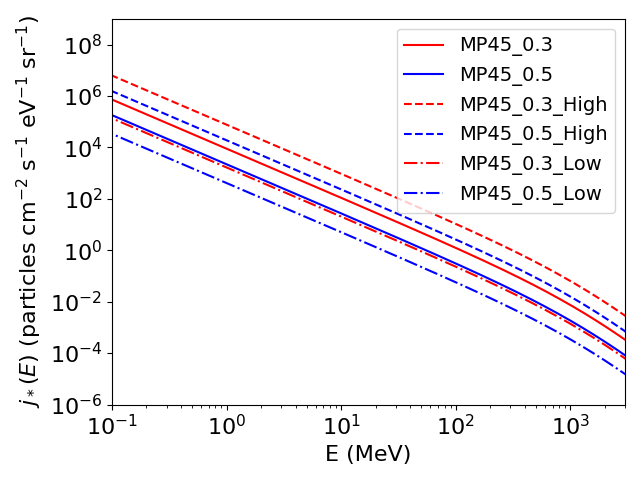}
\caption{CR proton spectra at the accretion shock at times $t=0.3$ Myr and $t=0.5$ Myr. Solid lines show the input spectrum using fiducial values of $f$, $\eta$, $R_*$, and $\dot{m}$, while the dashed and dashdot lines show the 'high' input CR spectra and 'low' CR input spectra. The slopes follow a power law with index $-1.9$ at lower energies.}
\label{fig:initial_spectrum}
\end{figure}

Figure \ref{fig:initial_spectrum} shows the initial spectra calculated for protons at the two evolution times. At lower energies, the spectra have a power-law behavior with index -1.9. The index in Equation \ref{equation:spectrum_power} is $\alpha\approx6$, so $j(p) \propto p^2 f(p) \propto p^{-4}$ \citep{amato_2015_CRs}. At lower energies, $p \propto \sqrt{E}$, so $j(E) \propto E^{-2}$. Above 938 MeV (the proton rest mass), $p \propto E$, so the spectrum turns over and eventually limits to $j(E) \propto E^{-4}$. 

\begin{table*} 
\centering
\begin{tabular}{ |c c c c c c c c c| } 
\hline
 Name & Time (Myr) & $M_{\rm core, 10^{4}} (M_{\astrosun})$  & $M_{*} (M_{\astrosun})$ & $R_{*} (R_{\astrosun})$ & $\dot{m} (M_{\astrosun}/\t{yr})$ & $L_{*} (L_{\astrosun})$ & $f$ & $\eta$ \\
 \hline
MP45\_0.3\ & 0.3 & 1.76 & 0.83 & 7.15 & $2.55\times 10^{-6}$ & 9.27 & 0.1 & $10^{-5}$ \\
MP45\_0.5\ & 0.5 & 0.79 & 1.07 & 6.83 & $4.39\times 10^{-7}$ & 2.15 & 0.1 & $10^{-5}$ \\
MP45\_0.3\_High & 0.3 & 1.76 & 0.83 & 7.15 & $2.55\times 10^{-5}$ & 92.68 & 0.9 & $10^{-4}$ \\
MP45\_0.3\_High & 0.5 & 0.79 & 1.07 & 6.83 & $4.39\times 10^{-6}$ & 21.53 & 0.9 & $10^{-4}$ \\
MP45\_0.3\_Low & 0.3 & 1.76 & 0.83 & 3.58 & $2.55\times 10^{-7}$ & 1.85 & 0.05 & $10^{-6}$ \\
MP45\_0.3\_Low & 0.5 & 0.79 & 1.07 & 3.42 & $4.39\times 10^{-8}$ & 0.43 & 0.05 & $10^{-6}$ \\
 \hline
\end{tabular}
\caption{Properties of the protostar that we use to construct the initial CR spectra. First three columns are the simulation name, time, and core mass. Remaining columns are protostellar mass, radius, accretion rate, and accretion luminosity, and the stellar filling fraction and shock efficiency. All runs assume a stellar magnetic field of $B_*=10$ G. The input CR spectrum for each setup is shown in Figure \ref{fig:initial_spectrum}}.
\label{table:star_parameters}
\end{table*}

We inject the CRs at the location of the protostar, which is located at  [X,Y,Z]=[0.00396, -0.00488, -0.00191] pc for the $t=0.3$ Myr grid and at [X,Y,Z]=[0.01584, -0.01579, -0.01428] pc for the $t=0.5$ Myr grid, shown by the white stars in Figure \ref{fig:bfield}. To calculate the energy of each particle, we divide the input energy spectrum into $N_E=50$ logarithmically spaced energy bins between $E_{\rm min}$ and $E_{\rm max}$, and sample the same number of initial particles $N_{PE}=1000$ from each bin. The total number of particles is $N=50 N_{PE}$. Running a higher number of particles produces less statistical error; but is more expensive. We discuss our choice of $N_{PE}$ in further detail and provide convergence tests in Appendix \ref{appendix_subsection:nparticles}. 

We compute the proton propagation in two stages. First, we use the highest resolution AMR grid and propagate the particles from $R_* \approx 0$ pc out to $R_{\rm start}=0.001$ pc, which is approximately the size of one grid cell in the lowest resolution grid. This gives us a modified spectrum $j_{\rm start}$ at $R_{\rm start}$. We include this step to capture the higher resolution magnetic field and density information close to the star where the gas densities are higher and may cause significant CR energy losses. We then use the lower resolution AMR grid to propagate the particles from $R_{\rm start}$ out to the edge of the protostellar core at $R_{\rm core}=0.1$ pc. We switch to using only this grid for most of the CR propagation to reduce the computational expense of the simulations and ensure that we run particles all the way to the edge of the core.  

For the first and second stages of the propagation, the magnitudes of $j_*$ and $j_{\rm start}$ respectively are captured in the weight function of each particle. Each particle begins with an energy $E_{i,\rm init}$. The weight of each particle is given by 
\begin{equation}
    \label{eq:weight}
    W_i=\frac{j_{\rm init}(E_{i,\rm init})4\pi R_{\rm init}^2 dE(E_{i, \rm init})}{N_{PE}}  \frac{\t{particles}}{s}.  
\end{equation}
where $R_{\rm init}=R_{*}$ and $j_{\rm init}=j_*$ for stage 1 and $R_{\rm init}=R_{\rm start}$ and $j_{\rm init}=j_{\rm start}$ for stage 2. Adopting this weight function allows us to use the same number of particles $N_{PE}$ in each bin and thereby reduce statistical uncertainties, while preserving the initial flux information at each energy. For the second stage runs, particles do not change direction or experience energy losses below $R_{\rm start}$, thereby preserving the information encoded in their weight function from $j_{\rm start}$. We note that this is not a perfect approximation, since it does not allow particles to propagate back inside $R_{\rm start}$ and lose energy.

\subsubsection{CR Propagation Parameters}
\label{subsubsection:crparameters}

The two problem-specific parameters that are not specified by the outputs of the \cite{offner_2017_impact} simulations are the scattering length (\ref{eq:lambda_sc}), particularly its dependance on the turbulence $\delta B(x)$, and the particle energy loss function (\ref{eq:loss}). The scattering length is calculated using Equation \ref{eq:lambda_sc}. The function $f(\beta_A)$ scales weakly with $\beta_A$, going as log($\beta_A$), so we follow \cite{fryer_2007_probing} and set it to a constant $f(\beta_A)= 5$. We assume the turbulence power spectrum to be scale-invariant, and Kolmogorov, which is described by a power-law index of $q=5/3$. Using these values the expression for the path length can be rewritten in terms of the gyroradius as \citep{fraschetti_2018_mottled}: 

\begin{equation}
    \label{eq:lambda_sc_2}
    \lambda_{\t{sc}}\approx 5.6L_c^{2/3}r_g^{1/3}/\sigma^2. 
\end{equation}

Equation \ref{eq:lambda_sc_2} shows that the path length is only $\propto r_g^{1/3}$, and so it is not a strong function of the coherent magnetic field strength or particle energy. However, it is a stronger function of the correlation length $L_c$ and the power of the turbulence $\sigma^2$. The value of $L_c$ is itself an uncertain parameter. Measurements of turbulence in molecular clouds have produced values of $L_c \sim 10^{-3}-10^{-1}$ pc \citep{houde_2011_bfields, houde_2016_bfields}; however, local outflow driving may produce smaller scale turbulence that is unresolved by detectors. Values of the correlation length that are this large give values for the path length that are larger than the size of our grid cells. In the \cite{offner_2017_impact} simulations, turbulence is followed self-consistently from the size of the outflow down to the size of a few grid cells ($\approx 0.001$ pc), but smaller scales are effectively under-turbulent due to numerical dissipation. As discussed in Section \ref{subsubsection:numerical}, our algorithm requires particles to resample their path length but not their direction when they enter a new grid cell; therefore, the path length should be sufficiently small that the particles are resampling their direction a few times in each cell. Given this, we adopt $L_c= 10^{-5}$ pc, which explores the possibility of small scale turbulence that may result from local outflow driving within dense cores and is not resolved by detectors or by the simulation. We also test $L_c= 10^{-4}$ pc, which is included in the Appendix. 

We also vary the relative power of the turbulence by exploring different values of $\sigma^2$. As in the case of $L_c$, the value of $\sigma^2$ is not well constrained for dense cores. We assume $\sigma^2$ is constant across the grid for each run (similar to \cite{fraschetti_2018_mottled}). \cite{houde_2009_bfields} measured the value of $\langle\delta B^2\rangle/\langle B_0^2\rangle$ in OMC-1 and found $\langle\delta B^2\rangle/\langle B_0^2\rangle=0.28$. In contrast, \cite{pineda_2020_turbulence} estimated $\delta B \approx 27 \mu \rm G$ inside the dense core Barnard 5, which is $\approx 5 \%$ of the previously estimated value of the static magnetic field of $B_0 \approx 500 \mu \rm G$ (corresponding to $\sigma^2 \approx 0.003$). We study three different values- $\sigma^2=0.01$, $\sigma^2=0.1$, and $\sigma^2=1.0$, as did \cite{fraschetti_2018_mottled}. This spans a wide range of possibilities for the turbulent power, from conditions where the turbulence is negligible ($\sigma^2=0.01$) to those where it is significant ($\sigma^2=1.0$). 

To model energy losses, we use the loss function from \cite{padovani_2009_cosmic} for protons in a molecular cloud. Between 100 keV and 1 GeV, energy losses are dominated by collisions with molecular hydrogen, exciting electronic excitations and inducing ionizations. At GeV and higher energies, energy losses are dominated by pion production, resulting in gamma radiation. 

\section{Results}
\label{section:results}

\subsection{CR Flux in Protostellar Cores}
\label{subsection:crflux}

The flux as a function of radius is computed as 
\begin{equation}
j(E_b, r)= \frac{1}{4\pi r^2 dE} \frac{1}{\overline{1/\mu_i}}\sum_{N_{\rm hit}(r)} \lp\frac{W_i}{\mu_i}\rp  \frac{\rm particles} {\rm s cm^2 eV},
\end{equation}
where the sum is over $N_{\rm hit} (r)$ particles that reach position $r$ in bin $E_b$, $\mu_i=\hat{n}\cdot \hat{v}$ is the cosine of the crossing angle of particle $i$, and $W_i$ is the weight of particle $i$. The subscript $b$ for the energy refers to the fact that the particle is in an energy bin $b$ and thus can have a range of values. Using the weight function given in Equation \ref{eq:weight}, the flux can be written:
\begin{equation}
\label{eq:radial_flux}
j(E_b, r)=\lp\frac{R_*}{r}\rp^2 \frac{1}{N_{PE}} \frac{1}{\overline{1/\mu_i}} \sum_{N_{\rm hit} (r)} \lp\frac{j_{\rm init}(E_{i,\rm init})}{\mu_i}\rp  \frac{\rm particles} {\rm s cm^2 eV}.
\end{equation}

\begin{figure*}
\centering
\includegraphics[width=0.98\linewidth]{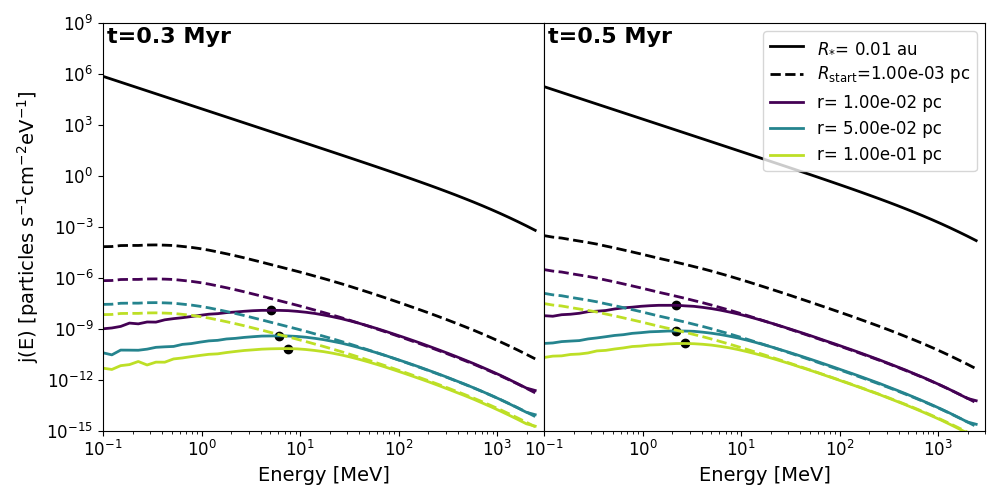} \caption{CR proton energy spectrum as a function of radius at $t=0.3$ Myr and $t=0.5$ Myr. Solid black lines show the unattenuated spectrum at $R_*$, while dashed black lines show the attenuated spectrum at $R_{\t{start}}=0.001$ pc. Colored lines show the spectrum at different distances from the star, with the free streaming solution from $R_{\rm start}$ plotted in dashed lines. Black dots show $E_{\rm turn} (r)$, which is the energy at which the spectrum is a maximum. Below $E_{\rm turn} (r)$, the spectrum turns over due to energy losses.}
\label{fig:t_radial}
\end{figure*}

We first compute the CR proton spectrum as a function of radius away from the protostar for our fiducial configuration, $\sigma^2=0$. Figure \ref{fig:t_radial} shows the attenuated CR proton spectra computed at $t=0.3$ Myr and $t=0.5$ Myr. For both times, the spectra show an overall decline with radius due to geometric attenuation and a turnover in the spectrum due to energy losses. The similarity of the spectra at higher energies to the solutions at $R_{\rm start}$, shown by the dashed black curves, implies that particles of these energies do not change energy significantly, as this would change the slope. Additionally, at higher energies the spectra follow the free streaming solution, shown by the colored dashed lines in the panels. The point at which the spectra turn over, $E_{\rm turn} (r)$, is roughly equal to the point at which they deviate from the free streaming solution. The similarity of the spectrum to the free streaming solution at energies above $E_{\rm turn} (r)$ implies that the geometry of the magnetic field does not change the particle travel directions significantly, because if the particles scattered significantly the flux at these energies would be higher than the free streaming solution. 

\begin{figure*}
\centering
\includegraphics[width=0.98\linewidth]{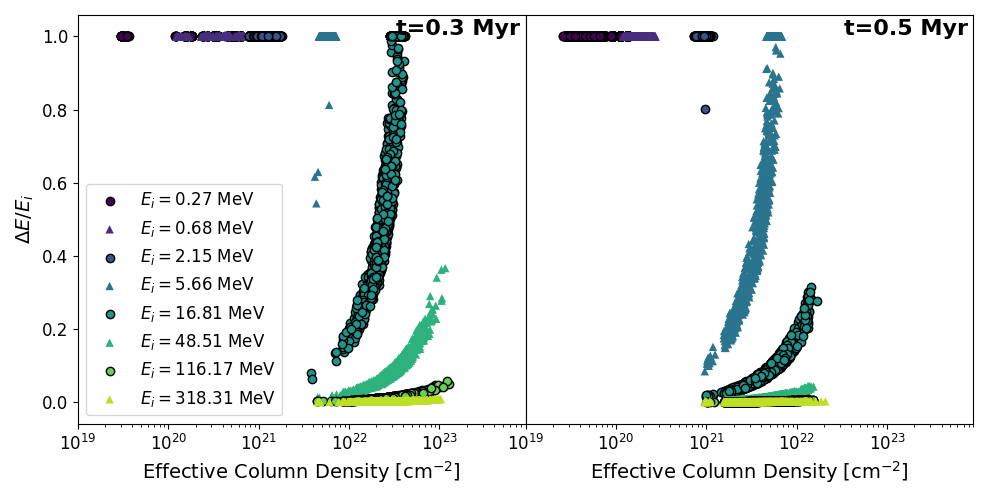}
\caption{Relative change in energy experienced by 1000 particles per initial energy for $t=0.3$ Myr and $t=0.5$  Myr as a function of effective column density they pass through as they propagate, starting at eight different energies.}
\label{fig:deltaE}
\end{figure*}

Figure \ref{fig:deltaE} shows the relative change in energy $\Delta E /E_i$ experienced by a subset of 1000 particles out to 0.1 pc as a function of the effective column density they pass through as they propagate for both times. At both times, the lower energy particles lose all their energy before traveling through very much of the core material while the higher energy particles can propagate through most of the gas without losing any energy. However, the particle behavior between the two grids differs at intermediate energies ($E_i=5.66$ MeV, $E_i=16.81$ MeV, and $E_i=48.51$ MeV) due to differences in the density and magnetic field. At $t=0.3$ Myr, the particles traveled through an average column density of $N \sim 10^{22}-10^{23} \t{cm}^{-2}$, while at $t=0.5$ Myr the particles traveled through a column density of $N \sim 10^{21}-10^{22} \t{cm}^{-2}$. As a result, in the $t=0.3$ Myr run particles with higher initial energies experienced more significant energy losses than at the $t=0.5$ Myr, and the value of $E_{\rm turn} (r)$ is slightly greater at $t=0.3$ Myr than at $t=0.5$ Myr.

\subsection{Impact of Outflow Geometry on CR Propagation}
\label{subsection:geometry}

During particle propagation, the code tallies the particle position as a function of solid angle in addition to radius. The flux in one angular bin is calculated as 
\begin{eqnarray}
\label{eq:angle_flux}
j(E_b, r, \theta_b, \phi_b)=\lp\frac{R_*}{r}\rp^2 \frac{N_{\theta}N_{\phi}}{N_{PE}}  \frac{1}{\overline{1/\mu_i}} \times \nonumber \\ 
\sum_{N_{\rm hit} (r, \theta_b, \phi_b)} \lp\frac{j_{\rm init}(E_{i,\rm init})}{\mu_i}\rp  \frac{\rm particles} {\rm s cm^2 eV},
\end{eqnarray}
where $N_{\rm hit} (r, \theta_b, \phi_b)$ is the number of particles that reached position $r$ in bin $(\theta_b, \phi_b)$ and the extra factors of $N_{\theta}$ and $N_{\phi}$ are the numbers of polar and azimuthal angle bins respectively and account for the crossing area of each bin relative to that of the entire sphere. The angular bins are discretized uniform in cos($\theta$) and $\phi$ using $N_{\theta}=40$ and $N_{\phi}=40$, giving each bin a solid angle of $\Omega=4\pi/(N_{\theta}N_{\phi})=7.85 \times 10^{-3}$ sr. The subscripts $b$ for the solid angle quantities refer to the fact that, similar to energy, particles can have a range of values of $\theta$ and $\phi$, and we do not add to the tally when $\theta_b$ or $\phi_b$ changes for a particle $i$. Since Equation \ref{eq:angle_flux} accounts for the area of each bin, it would give the same result for each angular bin as Equation \ref{eq:radial_flux} if particle propagation were isotropic, as has been assumed in previous studies. 

\begin{figure*}
\centering
\includegraphics[width=0.98\linewidth]{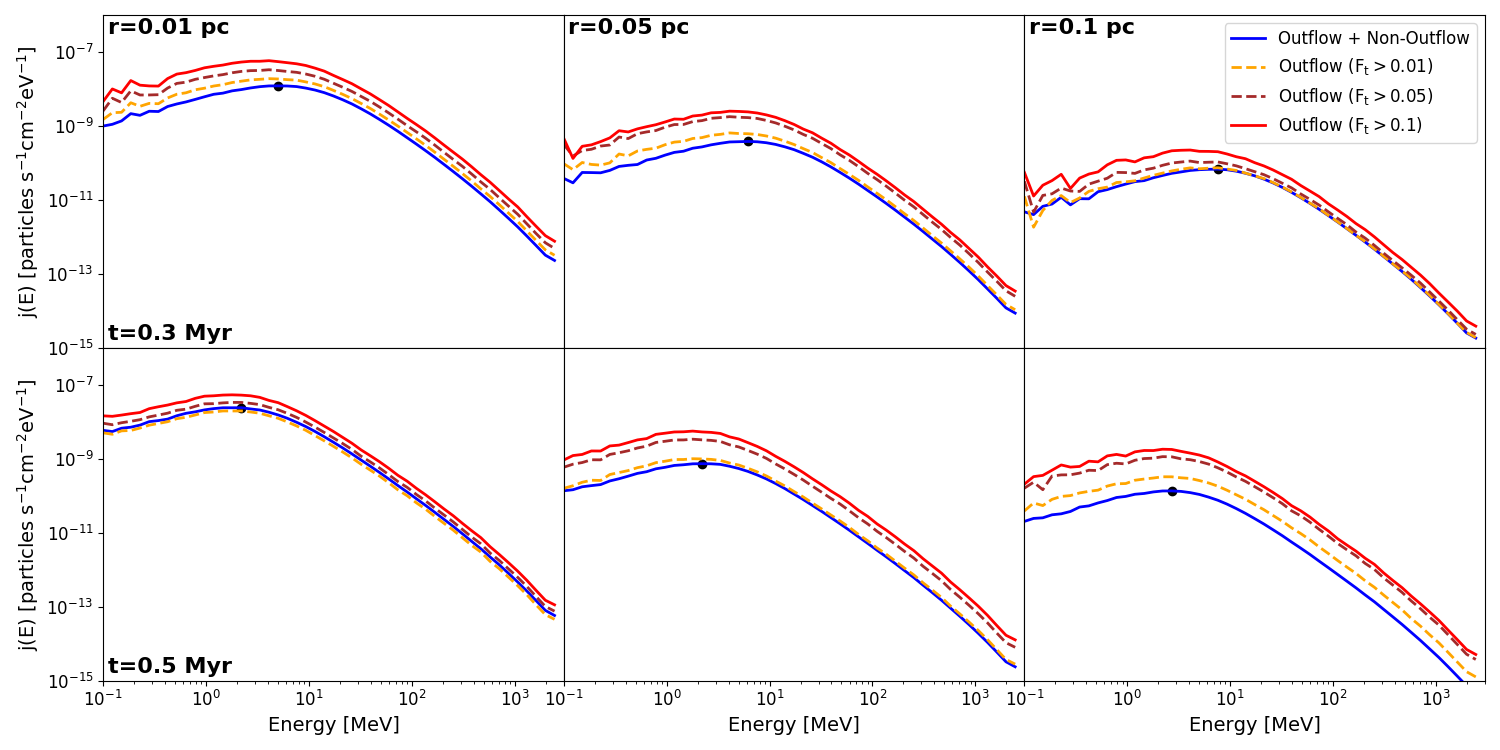}
\caption{Average flux for all spherical bins with $F_t>0.01$ (yellow), $F_t>0.05$ (red), and $F_t>0.1$ (brown) for $t=0.3$ Myr (top) and $t=0.5$ Myr (bottom) at $r=0.01$ pc, $r=0.05$ pc, and $r=0.1$ pc. Black dots show $E_{\rm turn} (r)$, which is the energy at which the radial spectrum is a maximum.}
\label{fig:flux_outflow}
\end{figure*}

To distinguish the location of the outflow,we compare the flux in different regions of the grid with different values of the tracer fraction $F_t$. We compute the tracer fraction in spherical coordinates and then categorize every spherical bin as being either in the outflow with $F_t>0.1$ or outside the outflow with $F_t<0.1$. We then compute the average values of $j(E_b, r, \theta_b, \phi_b)$ using different lower limit cutoffs of $F_t$. We denote these averages over select angular bins at a radius $r$ as $j(E_b, r)_{> F_t}$. Figure \ref{fig:flux_outflow} shows these results for three different radii- $r=0.01$ pc, $r=0.05$ pc, and $r=0.1$ pc. We define $j(E_b, r)_{> 0.1} = j(E_b, r)_{\rm out}$ as the average flux in the outflow region.

For both times, the flux in the outflow cavity is higher at all energies and distances than the average flux in the core. This is due to two possibilities. The first is that the magnetic field becomes aligned with the outflow region and can funnel particles into it, creating a higher flux inside the outflow region than outside. The second is that the gas densities inside the outflow are lower then the gas densities outside the outflow, decreasing energy losses in these regions. 

The prevalence of the magnetic field alignment can be deduced by comparing the shape of the average flux spectrum and the outflow flux spectrum above $E_{\rm turn} (r)$. At energies above $E_{\rm turn} (r)$, particles are not affected by higher density regions and are only affected by the magnetic field properties. This means that the magnetic field behavior can be deduced from the spectrum at these energies, and if the flux in the outflow is higher than average at these energies, there is a tendency for the magnetic field to be leading particles into the outflow. This can be seen in all six panels of Figure \ref{fig:flux_outflow}. 

The spatial properties of the gas density can be inferred by comparing the offset of the outflow flux from the average flux above and below $E_{\rm turn} (r)$. If the offset between the curves is higher below $E_{\rm turn} (r)$ than above $E_{\rm turn} (r)$, then there is a positive correlation between lower gas  densities and the outflow region. This behavior can be seen in Figure \ref{fig:flux_outflow} for all three panels of $t=0.3$ Myr (top row). For the three panels of $t=0.5$ Myr (bottom row), the spectrum is relatively even across all energies, indicating a minor correlation or no correlation; although there is the caveat that at large distances low energy particles are present mostly because they are repopulated by the higher energy particles losing energy, so in some sense they are 'following' what the higher energy particles are doing and it is harder to infer density information.

\begin{figure*}
\centering
\includegraphics[width=0.98\linewidth]{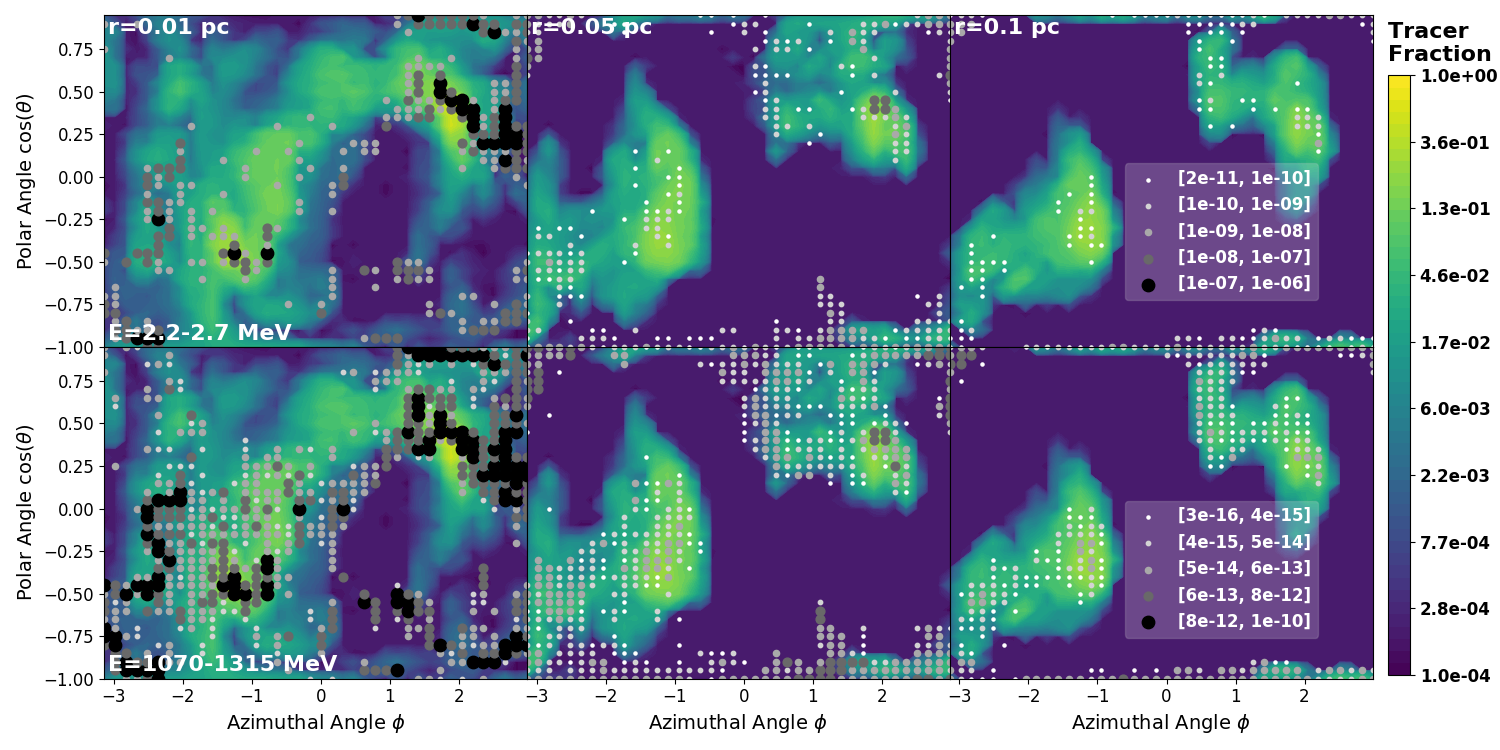}
\caption{Flux in particles $\rm s^{-1} cm^{-2} eV^{-1}$ for energy bins with $E_b=2.2-2.7$ MeV (top) and $E_b=1070-1315$ MeV (bottom) computed in different regions of the grid for $t=0.3$ Myr, at $r=0.01$ pc, $r=0.05$ pc, and $r=0.1$ pc. The legends indicate the range of flux values each dot represents, where larger, darker colored dots represent higher flux.
Regions without any dots have zero flux. Colorscale indicates the tracer fraction.}
\label{fig:flux_ebin_dots_t_03}
\end{figure*}

\begin{figure*}
\centering
\includegraphics[width=0.98\linewidth]{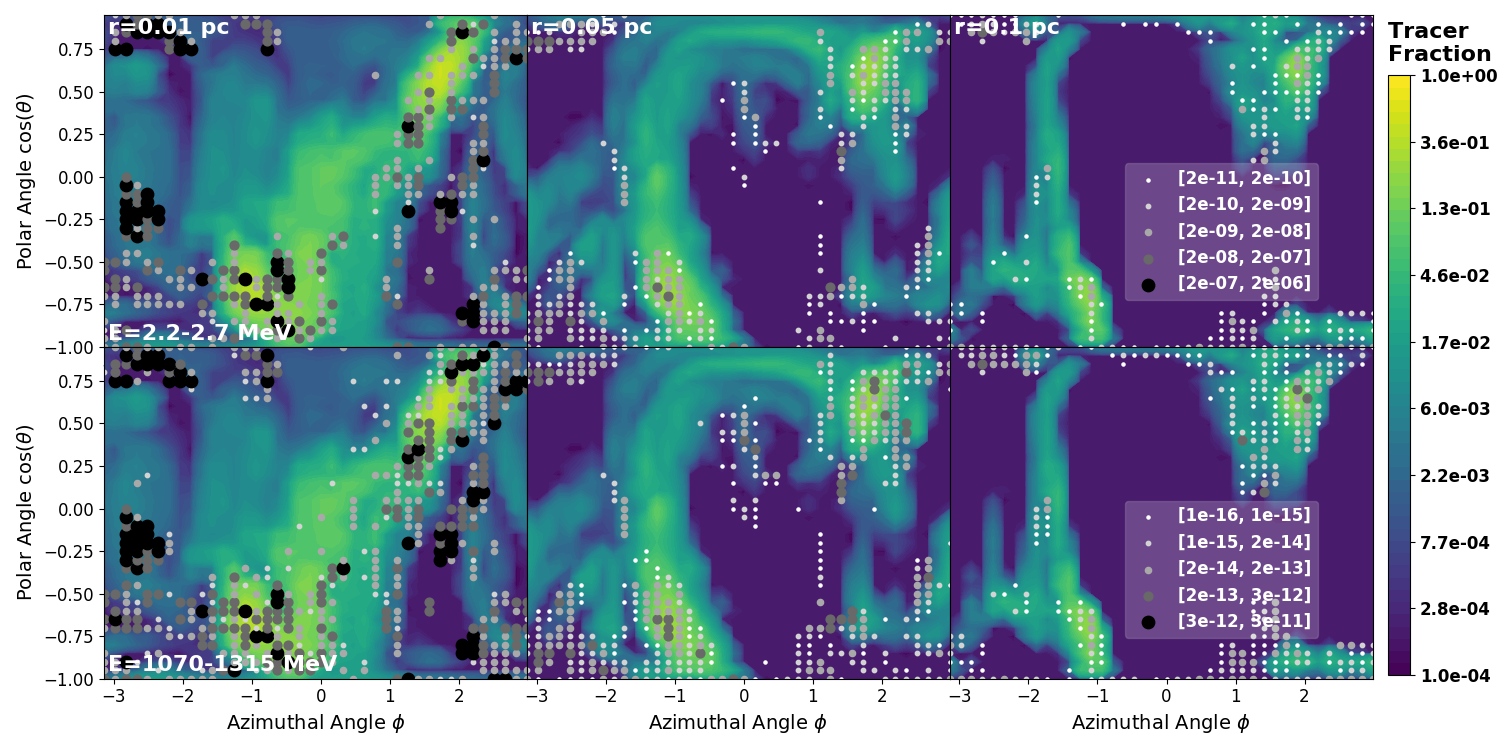}
\caption{Flux in particles $\rm s^{-1} cm^{-2} eV^{-1}$ for energy bins with $E_b=2.2-2.7$ MeV (top) and $E_b=1070-1315$ MeV (bottom) computed in different regions of the grid for $t=0.5$ Myr, at $r=0.01$ pc, $r=0.05$ pc, and $r=0.1$ pc. The legends indicate the range of flux values each dot represents, where larger, darker colored dots represent higher flux. Regions without any dots have zero flux. Colorscale indicates the tracer fraction.}
\label{fig:flux_ebin_dots_t_05}
\end{figure*}

Figures \ref{fig:flux_ebin_dots_t_03} and \ref{fig:flux_ebin_dots_t_05} show visual representations of the CR flux in a single energy bin computed at the two times. The size and color of the dots indicates the magnitude of the CR flux relative to the other regions of the panel, with blank areas indicating no CR flux in that area. These plots show the prevalence for particles to be more highly concentrated in the outflow, shown by the background colorscale, than in other regions of the grid. For example, for $E>E_{\rm turn} (r)$ for $t=0.3$ Myr (bottom row of plots in Figure \ref{fig:flux_ebin_dots_t_03}), all three panels show many regions with zero flux; however, the areas inside the regions with a higher tracer fraction for each plot are almost completely filled. This is what gives the outflow fluxes above the average at high energies shown in the top row of plots of Figure \ref{fig:flux_outflow}. The same effect is true at $E<E_{\rm turn} (r)$, depicted visually in the top row of plots of Figure \ref{fig:flux_ebin_dots_t_03}. 

These plots also show that the flux of high energy particles is more stochastically distributed than the flux of low energy particles. For example, at $r=0.01$ pc for $t=0.3$ Myr (left panels of Figure \ref{fig:flux_ebin_dots_t_03}), there are visibly more regions with a nonzero flux in the bottom plot depicting the flux for $E> E_{\rm turn} (r)$ than there are for the top plot depicting the flux for $E< E_{\rm turn} (r)$. This is because particles of all energies follow magnetic field lines, but particles of lower energies are kept out of certain regions due to higher gas density. The consequence of this is that all regions of the core with a nonzero flux of low energy particles also have a nonzero flux of high energy particles, but the converse is not always true.

\subsection{Cosmic Ray Ionization Rate}
\label{subsection:crir}

\begin{figure*}
\centering
\includegraphics[width=0.98\linewidth]{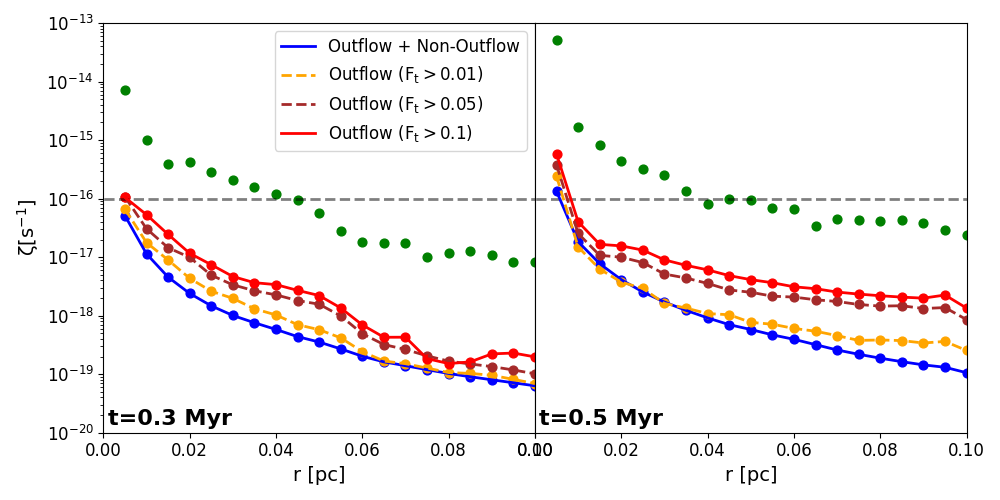}
\caption{Ionization rate as a function of radius for $t=0.3$ Myr and $t=0.5$ Myr. Blue lines show the average over all angular bins computed at a given radius, and the averages for $F_t>0.01$, $F_t>0.05$, and $F_t>0.1$ are shown in yellow, brown, and red respectively. Maximum values of any angular bin at a given radius are shown by the green points. The gray line indicates $\zeta=10^{-16} \rm s^{-1}$, which is the measured ionization rate for the Milky Way.}
\label{fig:crir}
\end{figure*}

Using the results for $j(E_b, r)$ shown in Figure \ref{fig:t_radial}, we compute the ionization rate. The ionization rate at a radius $r$ is calculated as:
\begin{equation}
    \label{eq:radial_crir}
    \zeta(r)=4\pi \int j(E_b,r) \sigma^{\t{ion}} (E_b) dE, 
\end{equation}
where $\sigma^{\rm ion}$ is the proton ionization cross section \citep{rudd_1985_cs, rudd_1991_cs, padovani_2009_cosmic}. We also compute the ionization rate as a function of solid angle. Similarly to Equation \ref{eq:radial_crir}, it is given by
\begin{equation}
    \label{eq:angle_crir}
    \zeta(r,\theta_b,\phi_b)= 4\pi \int j(E_b,r,\theta_b,\phi_b) \sigma^{\t{ion}} (E_b) dE.
\end{equation}
As with our results for the flux, we compute the average values of $\zeta(r,\theta_b,\phi_b)$ using angular bins with $F_t>0.01$, $F_t>0.05$, and $F_t>0.1$, and denote these averages $\zeta(r)_{>F_t}$. Figure \ref{fig:crir} shows the ionization rate at $t=0.3$ Myr and $t=0.5$ Myr.

For both times, the average ionization rate in the outflow is higher than the total average at all distances from the star. Furthermore, the offset between the average flux and the outflow flux is reflected in the offset of the average and outflow ionization rates. For example, as is seen in the bottom three panels of Figure \ref{fig:flux_outflow}, at $t=0.5$ Myr the discrepancy between the outflow flux and the average flux widens with increasing distance from the star. This is reflected in the right panel of Figure \ref{fig:crir}, in which the ionization in the outflow is further offset from the average ionization rate at further distances. This reflects particles being funneled into the outflow as they propagate outwards. 

The high gas densities in the core cause the average ionization rate to be lower than the fiducial value of $\zeta=10^{-16} \rm s^{-1}$, but some regions of the grid can have much higher ionization rates. For example, at $t=0.3$ Myr (left panel), the five innermost radii have average ionization rate values between $10^{-18}$ and $10^{-16} \rm s^{-1}$; but the peak ionization rates measured there are all between $10^{-16}$ and $10^{-14} \rm s^{-1}$. This highlights the importance of considering the geometry of the magnetic field and density in CR propagation calculations, as these can cause variations in the ionization rate greater than two orders of magnitude.

\begin{figure*}
\centering
\includegraphics[width=0.98\linewidth]{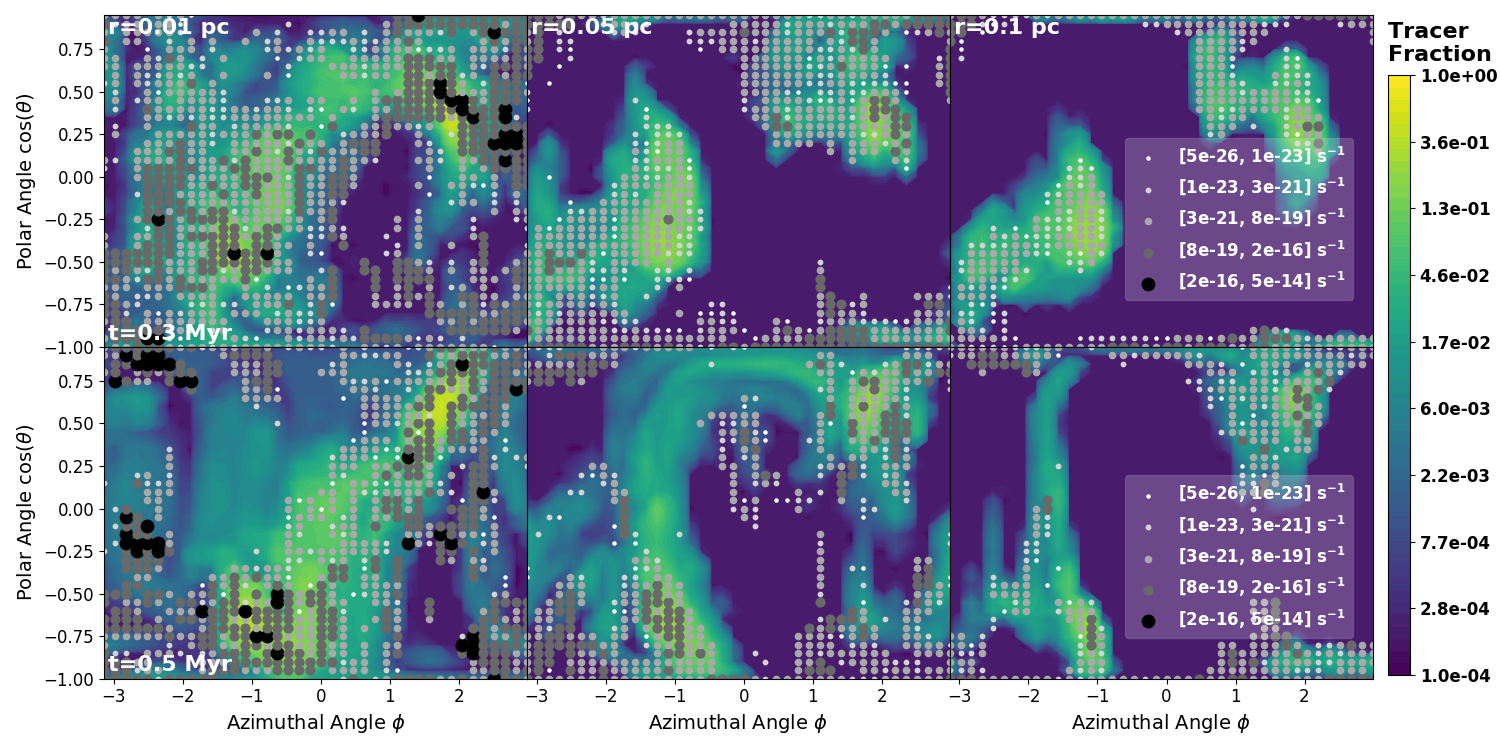}
\caption{Ionization rate for $t=0.3$ Myr (top) and $t=0.5$ Myr (bottom) computed in different regions of the grid at $r=0.01$ pc, $r=0.05$ pc, and $r=0.1$ pc. The magnitude of the ionization rate is indicated by the size and color of the dots which each represent a range of values, with larger darker colored dots indicating a higher ionization rate. Regions without any dots have no ionization. Colorscale indicates the tracer fraction.}
\label{fig:flux_crir_dots}
\end{figure*}

Figure \ref{fig:flux_crir_dots} shows the angular distribution of ionization rates for the two times at $r=0.01$ pc, $r=0.05$ pc, and $r=0.1$ pc. These plots have more dotted regions indicating a nonzero ionization rate than appear in Figures \ref{fig:flux_ebin_dots_t_03} and \ref{fig:flux_ebin_dots_t_05} because they include the influence of the flux at all fifty energy bins rather than a single bin as is plotted in the top and bottom rows of Figures \ref{fig:flux_ebin_dots_t_03} and \ref{fig:flux_ebin_dots_t_05}. However, many of the same regions that have a nonzero flux shown in Figures \ref{fig:flux_ebin_dots_t_03} and \ref{fig:flux_ebin_dots_t_05} are reflected by a higher ionization rate in Figure \ref{fig:flux_crir_dots}. For example, at $r=0.05$ pc for $t=0.3$ Myr (middle column of Figure \ref{fig:flux_ebin_dots_t_03}), the outflow regions have a higher average flux of both high and low energy particles than the averages for those energies outside the outflow. Similarly, the spatial distribution of the ionization rate for this case (top middle plot of Figure \ref{fig:flux_crir_dots}) shows a much higher average inside the outflow than outside, with all of the regions with no ionization occurring outside the outflow.

\subsection{Impact of Different Protostellar Properties}
\label{subsection:properties}

As discussed in Section \ref{subsection:initial}, several parameters affect the input CR spectrum. In particular, the accretion rate varies over several orders of magnitude and consequently has a large impact on the CR spectrum. Additionally, the radius of the star at any time is uncertain due to our accretion model. Finally, the other parameters of the CR spectrum model- the filling fraction, shock efficiency, and stellar magnetic field- are not well constrained observationally or theoretically. In this section, we assess the impact of different stellar assumptions on the resulting ionization rate. 

\begin{figure*}
\centering
\includegraphics[width=0.98\linewidth]{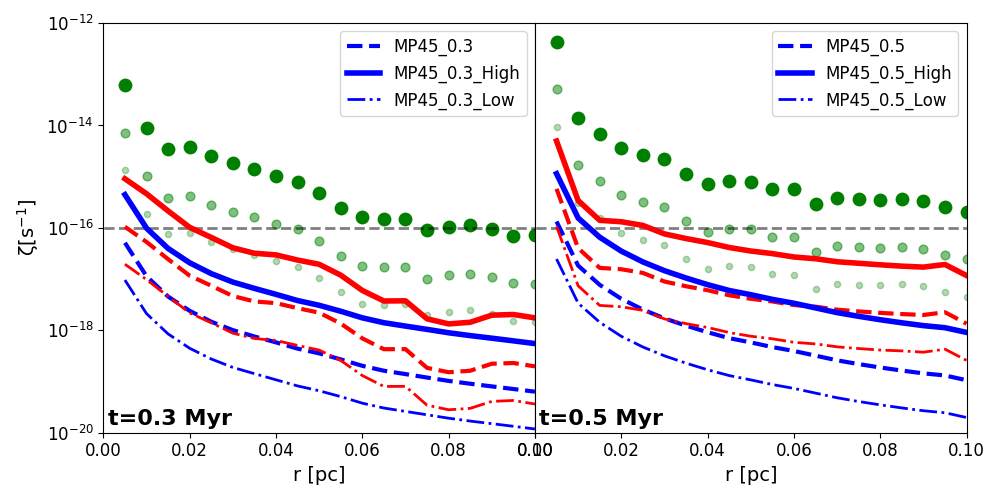}
\caption{Ionization rate as a function of radius for $t=0.3$ Myr and $t=0.5$ Myr for different input spectrum properties, indicated in Table \ref{table:star_parameters}. Blue lines show the averages over all angular bins computed at a given radius, averages for $F_t>0.1$ are shown in red, and the maximum values of any angular bin at a given radius are shown by the green points. Different linestyles indicate the evolution of the ionization rate assuming different CR spectra injected at the protostar, with the solid lines representing the highest initial spectra and the dash-dot lines representing the lowest initial spectra. The gray line indicates $\zeta=10^{-16} \rm s^{-1}$, which is the measured ionization rate for the Milky Way.}
\label{fig:crir_properties}
\end{figure*}

Figure \ref{fig:crir_properties} shows the ionization rate as a function of radius using the different input CR spectra indicated in Table \ref{table:star_parameters} and shown in Figure \ref{fig:initial_spectrum}. The high accretion rate models for both times have regions where the ionization rate is above the Milky Way value of $10^{-16} \rm s^{-1}$ throughout the core. The low accretion rate models did not have any regions where the ionization rate was above the Milky Way value except at r$<$0.01 pc. Although we examine the impact of the accretion rate at both times, we expect higher accretion on average in younger stars undergoing bursts of accretion. An important consequence of this is that we expect there to be a large variation in ionization over time, especially if the star is undergoing intense, frequent bursts.

One factor that we do not consider in these models is the coupling of the CRs to the gas in the accretion flow because it occurs on lengthscales smaller than the simulation grid resolution. As CRs travel through the accretion flow between the star and the inner disk they will experience energy losses \citep{offner_2019_disks}. The extent of the losses depends in part on the covering fraction of the accretion over the stellar surface, which is expected to be larger for higher accretion rates \citep{hartmann_2016_accretion}. Stars with low accretion rates may have significant enhancement of radionuclide production in the disk due to the enrichment from uncoupled CRs \citep{gaches_2020_radionuclide}. Although stars with high accretion rates have a higher unattenuated CR spectrum at the stellar surface due to the shock properties, the CRs may experience more energy losses close to the star compared to the low accretion case. \cite{offner_2019_disks} parametrize this uncertainty using a parameter $\epsilon$, which quantifies the interaction between the CRs and the accretion flow. A value  of $\epsilon=1$ indicates the accretion rate is constant and that all of the CRs traverse the full accretion column. Here, we essentially assume $\epsilon=0$ for all models; i.e., that CRs propagating through the core and outflow are not strongly coupled to the accretion streams and do not undergo energy losses. 

\subsection{Impact of Small Scale Magnetic Turbulence}
\label{subsection:turbulence}

\begin{figure*}
\centering
\includegraphics[width=0.98\linewidth]{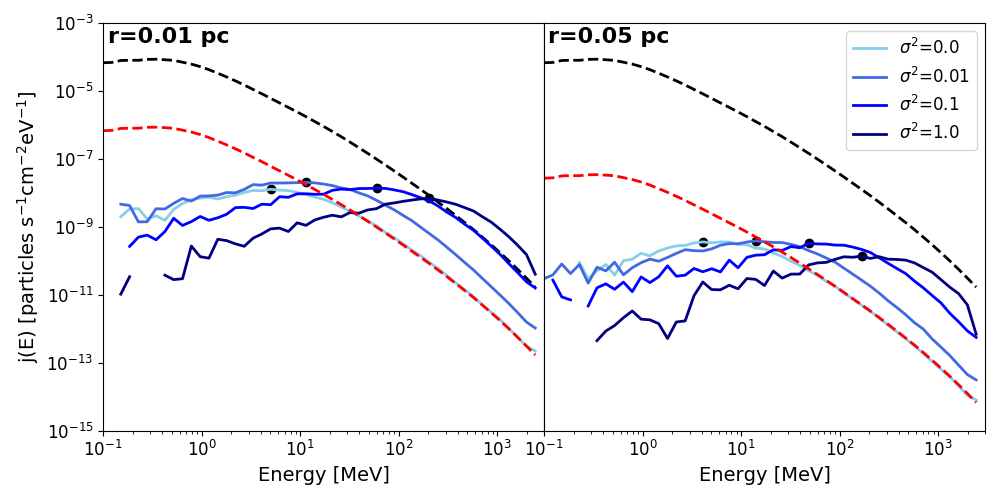}
\caption{CR energy spectrum at t=0.3 Myr for $r=0.01$ pc and $r=0.05$ pc using $\rm L_c=10^{-5}$ pc and four different values of $\sigma^2$. Black dashed lines show the attenuated spectrum at $R_{\rm start}=0.001$ pc. Dashed red lines show the free streaming solution at each radius. Black dots show $E_{\rm turn} (r)$, which is the energy at which the spectrum is a maximum.}
\label{fig:flux_radial_turbulence}
\end{figure*}

We computed the CR energy spectra as a function of radius using $\rm L_c=10^{-5}$ pc for different values of the turbulent power $\sigma^2$. Figure \ref{fig:flux_radial_turbulence} shows the results at two different radii for $t=0.3$ Myr. Increasing the amount of turbulence in the core from the fiducial value of $\sigma^2=0.0$ changes the shape of the spectrum and increases the value of $E_{\rm turn}$(r). This can be seen in both panels of Figure \ref{fig:flux_radial_turbulence}, in which the values of $E_{\rm turn}(r)$ are plotted as black points on each spectrum. The positive correlation in the flux with higher turbulence at energies greater than $\approx$ 500 MeV is caused by the increase in particle scattering. At low energies, energy losses dominate over scattering and more turbulence causes lower values of the flux. 

\begin{figure*}
\centering
\includegraphics[width=0.98\linewidth]{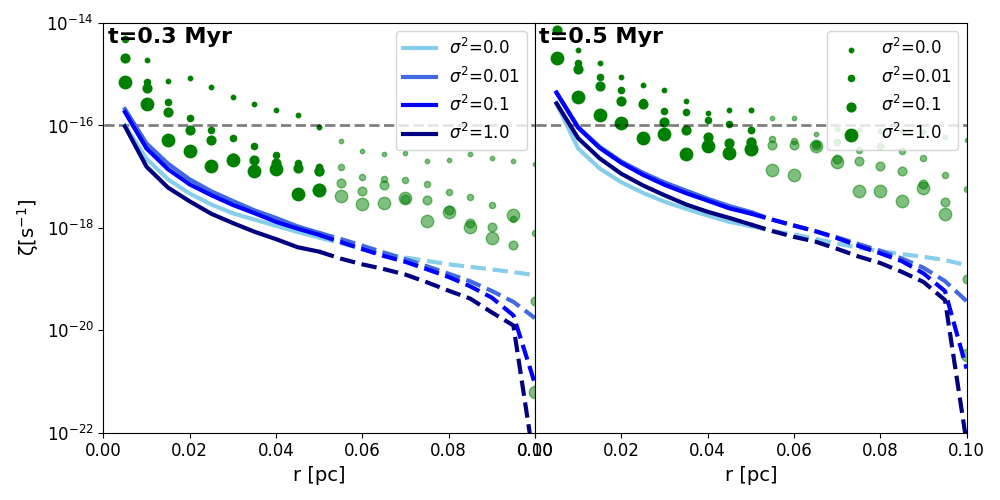}
\caption{Average ionization rate as a function of radius for $t=0.3$ Myr and $t=0.5$ Myr using $\rm L_c=10^{-5}$ pc. Turbulent power is indicated by the shade of blue. Maximum values of any angular bin at a given radius are shown by the green points, with larger dots symbolizing more turbulence. Dashed lines and translucent points at $r>0.05$ pc indicate the region where our results may be uncertain due to our simulation boundary conditions. The gray line indicates $\zeta=10^{-16} \rm s^{-1}$, which is the measured ionization rate for the Milky Way.}
\label{fig:flux_crir_turbulence_13_5}
\end{figure*}

Figure \ref{fig:flux_crir_turbulence_13_5} shows the average ionization rates for $t=0.3$ Myr and $t=0.5$ Myr. For most of the values of the turbulent power that we tested, there is a minimal change in the average ionization rates computed throughout the core. While stronger turbulence increases scattering and thus the flux of higher energy CRs, this contribution to the ionization rate is offset by the larger energy losses that attenuate the lower energy CRs. For example, for $t=0.3$ Myr, Figure \ref{fig:flux_radial_turbulence} shows that there is a noticeable difference in the spectra between $\sigma^2=0.01$ and $\sigma^2=0.1$, with $\sigma^2=0.1$ having a higher value of $E_{\rm turn}(r)$ for both distances from the core. However, the left panel of Figure \ref{fig:flux_crir_turbulence_13_5} demonstrates that the average ionization rates throughout the core for the two cases are indistinguishable. 

When there is a large enough amount of turbulence, the increase in the flux at high energies can no longer compensate for the energy losses at low energies and the average ionization rate decreases. The left panel of Figure \ref{fig:flux_crir_turbulence_13_5} displays this behavior for $\sigma^2=1.0$, where the value of the ionization rate for this case is lower at all distances from the core than for the lower values of $\sigma^2$. This effect is sensitive to density, as is demonstrated by the lack of this feature in the right panel of Figure \ref{fig:flux_crir_turbulence_13_5}.

\begin{figure*}
\centering
\includegraphics[width=0.98\linewidth]{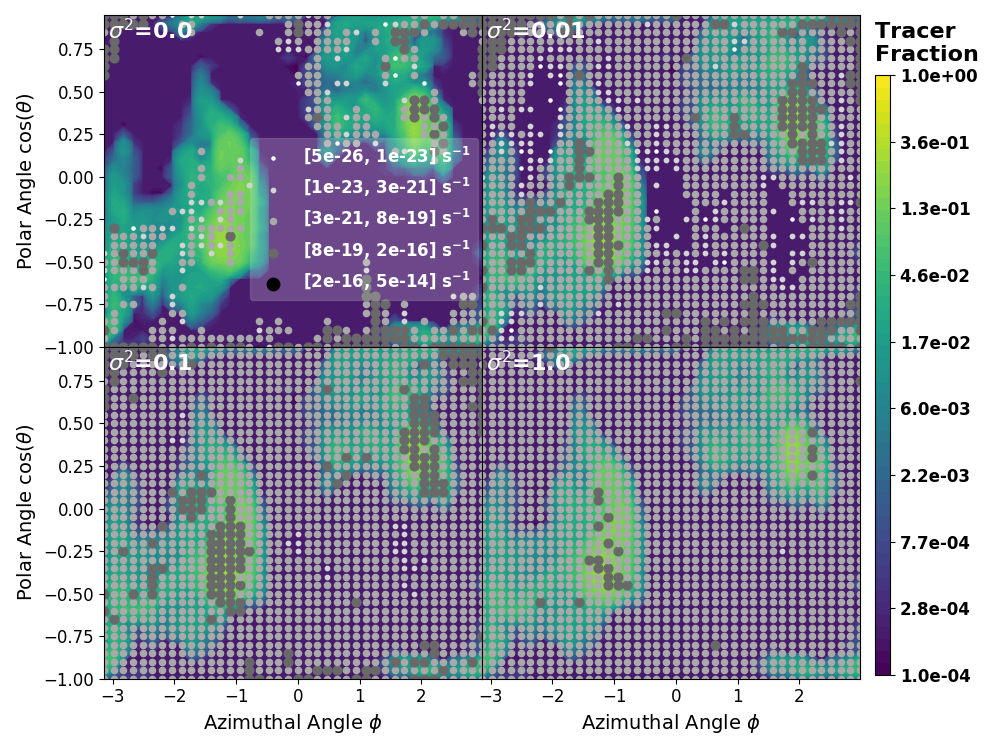}
\caption{Ionization rate for $t=0.3$ Myr using $\rm L_c=10^{-5}$ pc computed in different regions of the grid at $r=0.05$ pc for four different values of $\sigma^2$. The size and color of the dots indicates the magnitude of the ionization rate, with larger darker colored dots representing a higher ionization rate. Regions without any dots have no ionization. Colorscale indicates the tracer fraction.}
\label{fig:flux_crir_turbulence}
\end{figure*}

The distribution of CRs becomes more isotropic as the amount of turbulence increases.  This is because increasing the amount of turbulence reduces the influence of the bulk magnetic field, which introduces asymmetries by funneling particles into regions with a strong bulk magnetic field component. Figure \ref{fig:flux_crir_turbulence} shows the spatial distribution of ionization rates for $t=0.3$ Myr at $r=0.05$ pc for the four different values of $\sigma^2$. This figure demonstrates that even a small amount of turbulence changes the distribution of ionization rates; for $\sigma^2 = 0.01$ a larger fraction of the domain experiences non-zero ionization than for $\sigma^2 = 0.0$. This effect is even more pronounced for $\sigma^2=0.1$ (bottom left) and $\sigma^2=1.0$ (bottom right), where the entire grid at $r=0.05$ pc experiences a nonzero ionization rate.

As the ionization throughout the core becomes more isotropic, the maximum ionization rate declines.  For example, at $t=0.3$ Myr, Figure \ref{fig:flux_crir_turbulence_13_5} shows the maximum ionization rates for $\sigma^2=0.0$ are orders of magnitude higher than the maximum values for the nonzero values of $\sigma^2$. Increasing the amount of turbulence further decreases the maximum values, and the values for $\sigma^2=1.0$ are the lowest for the four values tested. Although turbulence increases the likelihood that a region of the core will have \textit{some} ionization, it decreases the chance of it having a particularly \textit{large} ionization.

Appendix \ref{appendix:turbulence_additional} shows results of the same calculations using $L_c=10^{-4}$ pc. We find that using a larger correlation length produces similar results.

\section{Discussion}
\label{section:discussion}

\subsection{Comparison With Observations}
\label{subsection:observations}

While the CR ionization rate from embedded protostars cannot be directly measured, the local ionization rate can be indirectly constrained through observations of the transition lines from various ions.  Several recent studies have used transition line measurements to place constraints on the ionization rates from different sources. In particular, $\rm N_2 H^+$ and $\rm HCO^+$ emission have been used to infer the ionization rates from the protocluster OMC-2 FIR-4 and the bow shock of the low-mass protostar L1157-mm, although these constraints have considerable uncertainties \citep{gaches_2019_mc1}.

OMC-2 FIR-4 is one of the closest known intermediate-mass protoclusters and has been observed in detail by the HIFI spectrometer aboard the \textit{Herschel} satellite \citep{ceccarelli_2014_omc}. It is located within the Orion Molecular Cloud (OMC) at a distance of 420 pc, has a total mass of approximately 30 $M_{\astrosun}$, and a total luminosity of $10^{3}$ $L_{\astrosun}$ \citep{kim_2008_omc, crimier_2009_omc, lopez_2013_omc2}. It contains three protostellar sources of approximately 1-10 $M_{\astrosun}$ each \citep{lopez_2013_omc2}. \cite{ceccarelli_2014_omc} used measurements of $\rm N_2H^+$ and $\rm HCO^+$ emission towards OMC-2 FIR-4 to estimate the ionization rate in the cluster. They assumed that most of the emission comes from the central source and treated it using a two-component model, with a warm inner region at 1250-3000 au ($6.1 \times 10^{-3}$-$1.5 \times 10^{-2}$  pc) from the source and a cold outer envelope at 3500-5000 au ($1.7 \times 10^{-2}$ - $2.4 \times 10^{-2}$ pc). They found that the inner and outer regions were well fit by ionization rates of $\approx 6\times 10^{-12} \rm s^{-1}$ and $4 \times 10^{-14} \rm s^{-1}$ respectively. \cite{fontani_2017_omc2} confirmed the estimate of $\zeta \approx 4 \times 10^{-14} \rm s^{-1}$ using measurements of $\rm HC_3N$ and $\rm HC_5N$, and \cite{favre_2018_omc} measured the same value using $\rm c-C_3H_2$. 

As can be seen in Table \ref{table:star_parameters}, none of our models have an accretion luminosity that is as high as that observed for OMC-2 FIR-4. The highest luminosity model that we have is the 'high' accretion rate model at $t=0.3$ Myr, which has a luminosity approximately ten times lower. Given this discrepancy, we do not expect to measure ionization rates as high as the observed value for OMC-2 FIR-4. The left panel of Figure \ref{fig:crir_properties} shows that for $r \approx 0.02$ pc, the average ionization rate we measure in the core was $\approx 2 \times 10^{-17} \rm s^{-1}$, roughly two orders of magnitude lower. Still, the measurements studied by \cite{ceccarelli_2014_omc} are not an azimuthal average and depend on the direction of the telescope relative to the outflow. The highest ionization rates we measure at $r \approx 0.02$ pc were $\approx 4 \times 10^{-15} \rm s^{-1}$, only a factor of ten lower than the \cite{ceccarelli_2014_omc} observations. Therefore, accounting for the luminosity difference, our model is consistent with their results. 

Another well studied source is the low mass Class 0 protostar L1157-mm, which has a luminosity of $\approx 4 L_{\astrosun}$ \citep{podio_2014_shock}; approximately equal to the luminosities of our fiducial models. The star drives the protostellar outflow L1157, which has a bright bow shock, L1157-B1, located at a distance of $\approx 0.1$ pc from the protostar \citep{podio_2014_shock}. \cite{podio_2014_shock} used emission measurements from L1157-B1 of $\rm N_2H^+$, $\rm HCO^+$ and other molecules to constrain the value of the ionization rate in the shock. They computed an ionization rate of $\zeta \approx 3 \times 10^{-16} \rm s^{-1}$. Figure \ref{fig:crir} shows that CRs can be directed into the outflow cavity, raising the average ionization rates in the outflow. Still, the maximum values we measure for our fiducial models at $r=0.1$ pc from the star are $\approx 10^{-17} \rm s^{-1}$, an order of magnitude lower than the \cite{podio_2014_shock} results. However, \cite{padovani_2016_protostars} showed that CRs can also be reaccelerated in the bow shocks of protostars, which may be a contributing source of CR acceleration in the L1157-B1 shock. Therefore, we do not consider our results completely inconsistent with the \cite{podio_2014_shock} results.  

In addition to predicting the radial evolution of the ionization rate, our models predict azimuthal variation, especially with respect to the location of the outflow cavity. Spatially resolved observations of molecular ions, such as HCO$^{+}$ and N$_2$H$^+$, which are sensitive to the CR ionization rate may provide evidence of non-uniform CR flux \citep{cleeves2015, seifert_2021_bfield}. The distribution and abundance of complex organic molecules, such as methanol (CH$_3$OH), acetaldehyde (CH$_3$CHO) and methyl formate (HCOOCH$_3$), which can be produced by CR heating of dust mantles \citep{ivlev2015,jimenez-serra2016,lopez-sepulcre2017,harju2020} may also point to elevated and/or anisotropic CR flux. However, obtaining high-resolution maps of these tracers toward protostellar cores is challenging, even with ALMA. In addition, external sources of ionization together with gas temperature and density variations may make it difficult to attribute any observed variation to CR anisotropies.

Since observations constrain the total ionization rate, it is also difficult to disentangle the relative contributions of  X-rays and CRs to the ionization. Young stars are strong X-ray emitters \citep[e.g.,][]{preibisch2005}. Hard X-ray photons produced through accretion and stellar flares
can provide a substantial source of ionization at intermediate densities \citep{glassgold1997,cleeves2015}. However, the radiation spectrum of protostellar sources, which have lower effective temperatures and larger radii than their T-Tauri counterparts \citep[e.g.,][]{hosokawa_2011_model}, are thought to have a softer X-ray spectrum and produce less ionizating radiation \citep{skinner1997}. Thus, we expect CRs are the dominant source of ionization in dense core envelopes and protostellar disks \citep{offner_2019_disks}.

Many previous studies have only considered the effect of CRs accelerated outside the molecular cloud on the cloud gas. While our calculations suggest that the elevated ionization rates observed in OMC-2 FIR-4 and L1157-B1 could be produced by local protostellar CRs, we note that a nearby supernova remnant (SNR) that efficiently accelerates CRs may have the same effect. SNRs are likely to be the main CR acceleration mechanism contributing to the Galactic CR (GCR) spectrum; however, they are probably not the only one. SNR remnants are unlikely to be able to accelerate particles up to PeV energies and cannot explain deviations observed in the isotopic ratios of the GCR spectrum from solar system values \citep{drury_2001_CRs, bykov_2020_acceleration}. While protostellar CRs do not explain these features, our calculations are able to explain the elevated ionization rates without assuming SNR acceleration, as a nearby SNR would likely produce other observational signals.

\subsection{Implications for the Chemistry of Protostellar Cores and Molecular Clouds}

Our results have implications for studies of the chemical abundances in protostellar cores and molecular clouds. These studies solve chemical networks to determine the evolution of the core or cloud chemistry and are used to link observations of the abundances to other properties. Very few studies consider internal sources of CR acceleration or even a spatially variable CR ionization rate \citep{gaches_2019_mc1}. 

Observations determining the abundances of deuterated molecules in protostellar cores are used to determine the core collapse history, which is important for determining the dynamical mechanism of core formation and evolution \citep{bacmann_2003_deuterium, flower_2006_deuterium, caselli_2008_deuterium, emprechtinger_2009_deuterium, fontani_2011_deuterium, kong_2015_deuterium, kong_2016_deuterium}. The deuterium fraction $D_{\rm frac}^{\rm N_2 H^+}$=[$\rm N_2 D^+$]/[$\rm N_2 H^+$] has been calculated using transition line measurements for a number of protostellar cores, with typical values of $\approx 0.1-1$, and used to place constraints on the ages of the cores \citep{crapsi_2005_deuterium, emprechtinger_2009_deuterium, pagani_2009_deuterium, pagani_2013_deuterium, kong_2016_deuterium}. However, these studies showed that variations in the value assumed for the CR ionization rate change the time evolution of the formation of $\rm N_2 D^+$  and the final equilibrium value of $D_{\rm frac}^{\rm N_2 H^+}$. Another observational indicator of the core age is the ratio of ortho- to para- abundances of deuterated molecules such as $\rm H_2 D^+$, but this also depends on the value assumed for the CR ionization rate \citep{caselli_2008_deuterium, brunken_2014_deuterium}. 

Several studies found variations in the core evolution when they varied the CR ionization rate from $\approx 10^{-18}- 10^{-15} \rm s^{-1}$ \citep{crapsi_2005_deuterium, pagani_2009_deuterium, fontani_2011_deuterium, pagani_2013_deuterium, brunken_2014_deuterium, kong_2015_deuterium, kong_2016_deuterium}. Figures \ref{fig:crir} and \ref{fig:crir_properties} show that the ionization rates throughout the core are not uniform and span this range, with higher values in certain regions such as the outflow. This suggests that the abundance depends on the viewing angle and measuring a single value of the deuterium abundance towards a protostellar core will not place a tight constraint on the age of the core. Additionally, some studies used observations of deuterium molecules to place upper limits on ionization rates in cores. \cite{kong_2015_deuterium} concluded that to reproduce the deuterium abundance observations in typical cores an ionization rate of $\zeta<10^{-16} \rm s^{-1}$ is needed. Figures \ref{fig:crir} and \ref{fig:crir_properties} show that there are regions of some protostellar cores with higher ionization rates than this, which indicates that the deuterium fraction may vary throughout the core.

Similar methods are used to determine how the CR ionization rate impacts the formation of neutral organic molecules in cores. \cite{shingledecker_2018_tmc1} calculated the abundances of HOCO, $\rm NO_2$, $\rm HC_2O$, and $\rm HCOOCH_3$ in a protostellar core, varying the ionization rate between $10^{-17}$ and $10^{-14} \rm s^{-1}$. They found an anti-correlation between CR ionization rate and neutral organic molecule abundance as the gas became dominated by ions. Figure \ref{fig:crir} shows that because of the heterogeneous distribution of ionization rates in the core, the maximum ionization rates throughout most of the core fall in the range $\zeta > 10^{-17} \rm s^{-1}$. This suggests that some regions of protostellar cores have reduced abundances of neutral organic molecules. 

In molecular clouds, the ionization rate affects the abundances of CO and CI, which are used to estimate the mass of $\rm H_2$ gas in the cloud \citep{papadopoulos_2004_C1, bollato_2013_CO, bisbas_2015_chemistry}. In order to use these transition lines as mass estimators, the relative abundances of $[\rm CO]/[\rm H_2]$ or $[\rm CI]/[\rm H_2]$ must be known. Numerical studies have calculated how these abundance ratios are affected by different values of the CR ionization rate in molecular clouds; however, they mostly assumed a constant CR ionization rate of $10^{-17} \rm s^{-1}$ throughout the cloud and examined the effect of varying it \citep{glover_2012_chemistry, offner_2013_chemistry, offner_2014_chemistry, glover_2015_chemistry_c1, glover_2016_chemistry_c1, bisbas_2017_chemistry}. An increase in the CR ionization rate causes a decrease in the abundance of CO and an increase in the abundance of CI, and at ionization rates 2-3 orders of magnitude greater than $10^{-17} \rm s^{-1}$ CO is almost completely depleted and CI is the dominant gas tracer \citep{glover_2012_chemistry, bisbas_2017_chemistry, gaches_2019_mc2}. 

\cite{gaches_2019_mc1, gaches_2019_mc2} found that including internal sources of CR acceleration from protostellar accretion shocks accurately reproduced the observed ionization rates and abundances of $[\rm CO]/[\rm H_2]$ and $[\rm CI]/[\rm H_2]$ in molecular clouds. Figures \ref{fig:crir} and \ref{fig:crir_properties} show that in regions of cores near the outflow and throughout the core in higher accretion sources the CR ionization rate is comparable to the value frequently assumed in the literature of $10^{-17} \rm s^{-1}$. Consequently, protostellar sources of CR acceleration will impact the abundances of CO and CI near the sources in molecular clouds. Studies using more realistic cloud morphologies and star formation distributions will calculate more accurate ionization rates in the clouds. 

Finally, we note that the CR flux in molecular clouds also affects the cloud temperature, as some of the CR energy goes into heating the gas. Consequently, clouds with significant CR ionization are also warmer by $\approx$ 30-50 K \citep{bisbas_2017_chemistry, gaches_2018_exploration}. Higher temperatures increase the thermal pressure of the gas, which may inhibit star formation \citep{mckee_2007_sf}. 

\subsection{Comparison With Prior Models}
\label{subsection:comparison}

Several studies have used analytical models to predict the CR ionization rate in protostellar cores. \cite{gaches_2018_exploration} calculated the CR attenuation and resulting ionization rates in protostellar cores using a simple radial density profile. They calculated the ionization for free streaming propagation, i.e. assuming the CRs are not coupled to a magnetic field, and for diffusive CR propagation, which corresponds to a turbulent magnetic field. Their fiducial model considered a 0.5 $M_{\astrosun}$ star and accretion rate of $\approx 1.4 \times 10^{-5} M_{\astrosun}/\rm yr$ (see their Equation 4), which is similar to our $\rm MP45\_0.3\_High$ model. The \citet{offner_2017_impact} simulations we use as inputs assume an initially uniform field in the $z$ direction, which imposes a degree of order even with the gas turbulence, so free streaming is a reasonable approximation for our calculations with $\sigma^2=0.0$. For this model they computed $\zeta \approx 10^{-14} \rm s^{-1}$ at $N\approx10^{21} \rm cm^{-2}$, which drops to $\zeta \approx 10^{-19} \rm s^{-1}$ at the edge of the core where $N \approx 10^{23} \rm cm^{-2}$ (see their Figure 11). As can be seen in Figure \ref{fig:crir_properties}, the average values we measure are between $ \approx 10^{-19} \rm s^{-1}$ and $10^{-16} \rm s^{-1}$, so our results are roughly consistent with those computed by \cite{gaches_2018_exploration}. However, Figure \ref{fig:crir_properties} also demonstrates that the CR ionization rate can vary substantially at a given radius and column density when asymmetries due to the magnetic field, gas structure and outflow are considered.

\cite{silsbee_2019_model} derived analytical solutions for the attenuation of the CR ionization rate including energy losses as a function of effective column density. Their solutions assume a bulk magnetic field $\boldsymbol{B_0}(x)$ and a turbulent magnetic field $\boldsymbol{\delta B}(x)$. The inputs of their model include the properties of the initial CR spectrum and loss function. Appendix \ref{appendix_subsection:tests_losses} provides further details of their model, modified to account for the spherical geometry of our problem.
We compare their analytic model to our Monte Carlo calculations. We compute the average radial density profile for both of our grids and derive what the \cite{silsbee_2019_model} solutions for the ionization rate would be given our initial spectra. We also perform calculations using this radially averaged density, which should give the same results. We expect these curves to be similar, but not necessarily equal to, the average ionization rates we compute for our grids shown by the curves in Figure \ref{fig:crir}.  

\begin{figure*}
\centering
\includegraphics[width=0.98\linewidth]{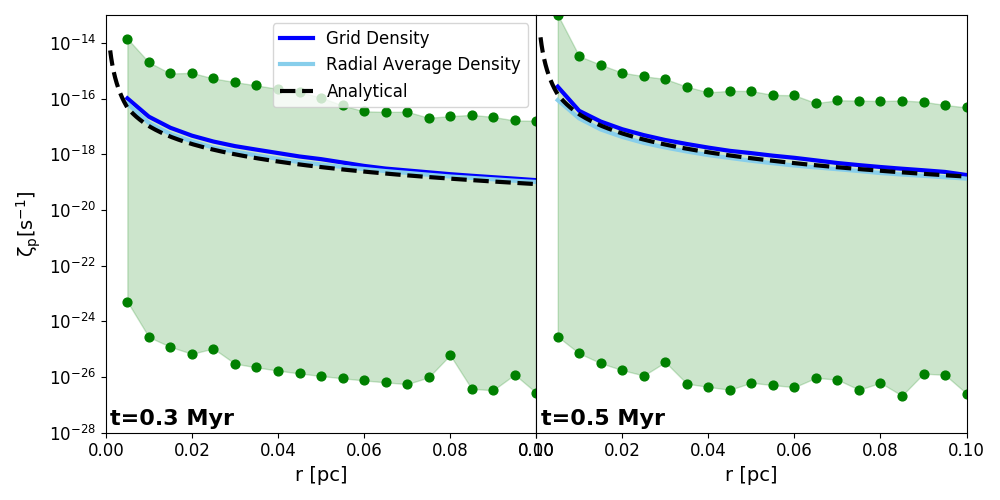}
\caption{
Average ionization rate as a function of radius for $t=0.3$ Myr and $t=0.5$ Myr. Dark blue curves use the densities in our grids. Light blue curves show numerical results using the radially averaged density profile in each grid. Black curves show the \cite{silsbee_2019_model} model solutions using this density profile and our input loss function and CR spectra. Green points show the maximum and minimum values of any angular bin at a given radius.
}
\label{fig:flux_crir_lines_comparison}
\end{figure*}

Figure \ref{fig:flux_crir_lines_comparison} shows the results of this test. As can be seen, the solutions are within a factor of a few of each other. However, the green points demonstrate that at each radius we compute a wide range of ionization rate values that are not captured by the \cite{silsbee_2019_model} model or by other studies which take a radial density profile. The ranges of ionization rates that we compute at each radius, shown by the green shading, are all greater than five orders of magnitude. This test shows that, while our calculations give similar average results as other models, they provide a more nuanced view of the ionization rates in different regions around the star.

\subsection{Caveats}
\label{subsection:caveats}

Our simulations make some approximations that may impact our results. For one thing, the properties of the protostellar disk are not resolved. To account for CR energy losses in the innermost regions of the core, we use the maximum resolution grid with $\Delta x_{\rm AMR}=26$ au to compute the radial attenuation of the CR spectrum out to $R_{\rm start}=0.001$ pc. This resolution does not resolve the circumstellar disk density and magnetic field. CR propagation in the disk may cause a non-radially symmetric CR spectrum at $R_{\rm start}$, with lower fluxes of CRs in regions near the disk due to energy losses inside the disk \citep{offner_2019_disks}. This could affect the angular distribution of ionization rates shown in Figure \ref{fig:flux_crir_dots}, especially at smaller radii, and cause lower ionization rates throughout the core.

As discussed in Section \ref{subsubsection:bfield}, our treatment of the two component magnetic field is less accurate than resolving the turbulence as some other studies do \citep[e.g.,][]{giacolone_1999_turbulence, fraschetti_2018_mottled}. Our method is less computationally expensive, but it is an approximation  of the behavior of particles in the presence of magnetic fields. Using the path length to sample the particle step size is a reasonable approximation over solving the full Lorentz equation, as the gyroradius is orders of magnitude smaller than the path length; however, by not resolving the turbulence, we are forced to choose the direction of travel statistically. Figures \ref{fig:flux_radial_turbulence}, \ref{fig:flux_crir_turbulence_13_5}, and \ref{fig:flux_crir_turbulence} show the qualitative effects that low and high amounts of turbulence can have on particle propagation and ionization throughout the core. Furthermore, as described in Appendix \ref{appendix:tests}, our code reproduces the correct solution in the free streaming and diffusion limits. Further studies are needed will show how well the quantitative results approximate the full numerical treatment. 

Additionally, our treatment of turbulence only accounts for spatial diffusion of the CRs. Momentum diffusion is included in many CR propagation codes and causes CR reacceleration \citep{strong_1998_galprop}. If the gas is strongly supersonic ($M_s>>1$) and super-Alfv\'enic ($M_{\rm A}>>1$) then it may be able to reaccelerate CRs significantly to relativistic energies. For both of our grids the radially averaged values of the sonic and Alfv\'enic Mach numbers had upper limits of $M_s<2$ and $M_{\rm A}<0.7$ for $r>0.02$ pc respectively, and less then 10 $\%$ of the gas in the core was both supersonic and super-Alfv\'enic. Therefore, we do not expect there to be efficient reacceleration of CRs in most of the core.

Our model also does not account for the effects of CR energy deposition, which may modify the turbulence in the gas. To examine how this would impact the outcome of the gas in the {\sc orion} calculations, we computed the radially averaged CR energy density and gas kinetic energy density throughout the core. We found that $\epsilon_{\rm KE}/\epsilon_{\rm CR} \approx 10-10^{5}$, which indicates that the energy budget of the CRs is not high enough to significantly alter the outcome of the gas. However, it is large enough to damp out the smallest scales of the turbulent energy, which may develop at lower scales than the resolution of the \cite{offner_2017_impact} simulations. Assuming that the turbulence roughly follows a Kolmogorov power spectrum ($ \epsilon_{\rm KE}(k) \propto k^{-5/3}$) with a numerical viscosity that is roughly the size of a grid cell, the CR energy would damp the turbulent energy at $\approx 1/100$th of the cell length. This provides reassurance that the viscosity set by the grid scale of the MHD simulations is only a couple orders of magnitude higher than the true viscosity set by CRs, and small scale turbulence is unlikely to significantly alter the evolution of the magnetic field in the MHD simulations.

Additionally, our code assumes time independence. Under this assumption, the initial CR spectrum at the accretion shock is the same for all times, and the speed at which the particles of different energies propagate does not effect our results. Clearly, this assumption is not true, as protostars undergo changes in their mass, radius, and accretion rate, which affect the initial CR spectrum. However, the timescales for CR propagation are much less than that for accretion. The time for a particle traveling at speed $c/100-c$ to cross a grid cell of 0.001 pc is $\approx 0.003-0.3$ years, or  $\approx$ 1-100 days. While small flickering in the accretion rate can occur on the timescales of days, episodic accretion is characterized by variations that occur on timescales of months to years \citep{audard_2014_ppvi}. This timescale is also less than the dynamical timescale of the gas to cross one cell, which is approximately a year. Therefore time independence is a reasonable assumption for our calculations.

Finally, our results are limited by the boundary conditions we impose for our grid. We discuss this further in Appendix \ref{appendix_subsection:boundary_conditions}. 

\section{Conclusions}

In this work, we use a Monte Carlo cosmic ray transport code to investigate the propagation of cosmic rays accelerated by protostellar accretion shocks in protostellar cores. In order to account for realistic physical asymmetries, we post-process simulations of turbulent, magnetized protostellar cores, which include protostellar outflows. The turbulence is modeled using a statistical method that samples the particle travel direction and step size as a function of the turbulent power, parametrized by $\sigma^2=(\delta B/B_0)^2$. We calculate the ionization rate as a function of both radius and solid angle around the star. We explore the effect of varying the accretion properties and amount of turbulence in the core on the CR propagation and resulting ionization rates.  

The CR flux is non-uniform when there is negligible sub-grid turbulence in the core and is focused in the direction of the outflow, which is similar to the `flashlight' effect observed for photons \citep{yorke_2002_stars, cunningham_2011_outflows}. This creates elevated CR ionization rates in the outflow region. We find that at a given distance from the star the average ionization rate in the outflow can be up to an order of magnitude higher than the ionization rate averaged over all angular bins at that radius. The maximum ionization rates calculated at a given radius can be up to two orders of magnitude or higher than the average at that radius. 

Our results are sensitive to the accretion rate and other stellar parameters. A two order of magnitude range in the accretion rate causes differences in the ionization rates in the core by up to two orders of magnitude. In contrast to the stellar parameters, varying $\sigma^2$ by two orders of magnitude increases the uniformity of the CR distribution and changes the shape of the spectrum but does not have a significant effect on the resulting ionization rates. The increased scattering of the CR particles with higher turbulence enhances the flux at high energies and decreases it at low energies, offsetting the net impact on the ionization rate. 

For our fiducial calculations, the radially averaged ionization rates throughout the core are not as high as the measured value for the Milky Way of $\zeta \approx 10^{-16} \rm s^{-1}$, due to attenuation by the dense gas. However, within $r \sim 0.04$ pc, the maximum ionization rates exceed the measured Milky Way value. Throughout most of the core, the maximum ionization rates exceed $10^{-17} \rm s^{-1}$, which is the value often assumed in astrochemistry studies. 

In summary, our calculations suggest protostellar cores with low levels of turbulence have non-uniform CR fluxes that produce a broad range of ionization rates. This has important implications for the chemistry of molecular clouds, protostellar cores, and accretion disks. We conclude that accurate models of CR propagation must take into account the geometry of the magnetic field and gas density.

\appendix

\section{Resolution Limitations}
\label{appendix:uncertainties}

\subsection{Grid Boundary Conditions}
\label{appendix_subsection:boundary_conditions}

Our results are limited by the boundary conditions we impose for our grid. In this section we investigate the impact of the domain size on our results. Figures \ref{fig:flux_crir_turbulence_13_5} and \ref{fig:flux_crir_turbulence_14_5} show the CR ionization rate out to 0.1 pc (roughly the size of the core); however, this distance is close to the edge of the grid, so we stop tracking CR propagation once the CRs reach this radius. Consequently, they cannot scatter back and increase the ionization as they would if the grid were much larger. We can use results calculated at a smaller radius to evaluate how accurate our results are with this caveat. For each value of $\sigma^2$ and $L_c = 10^{-5}$ pc, we compute the ionization rate out to $r_{\rm test}=0.025$ pc with three different values for the maximum radius of propagation $r_{\rm bound}$. We test $r_{\rm bound}=r_{\rm test}=0.025$ pc, $r_{\rm bound}=2r_{\rm test}=0.05$ pc, and $r_{\rm bound}=4r_{\rm test}=0.1$ pc. The tests with $r_{\rm bound}=r_{\rm test}$ are analogous to the results we present in Figures \ref{fig:crir} and \ref{fig:flux_crir_turbulence_13_5}, while the other two calculations offer insights into how the chosen grid boundary impacts the ionization rates.

\begin{figure*}
\centering
\includegraphics[width=0.98\linewidth]{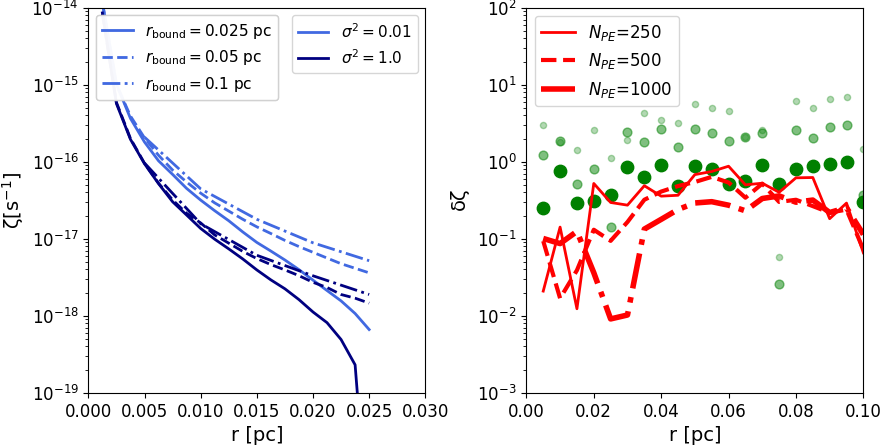}
\caption{Left: Ionization rate for three different values of $r_{\rm bound}$, using $L_c=10^{-5}$ pc at $t=0.3$ Myr. Right: Relative error in the outflow and maximum ionization rates, where the reference value is $N_{PE}=2\times 10^{3}$ particles. The model adopts the simulation at $t=0.3$ Myr, using $L_c=10^{-5}$ pc, and $\sigma^2=1.0$.}
\label{fig:crir_plots_uncertainties}
\end{figure*}

Figure \ref{fig:crir_plots_uncertainties} shows the results for $\sigma^2=0.01$ and $\sigma^2=1.0$. For each value of $\sigma^2$ the curves are within a factor of a few of each other out to $r \approx r_{\rm test}/2= 0.0125$ pc. Past this radius, the curves with $r_{\rm bound}=r_{\rm test}$ start to decline further, but the discrepancy is still less than an order of magnitude at $r \approx 4/5 r_{\rm test}=0.02$ pc. From these results, we conclude that the results in Figures \ref{fig:flux_crir_turbulence_13_5} and \ref{fig:flux_crir_turbulence_14_5} are accurate to within a factor of a few out to $r=0.05$ pc, and we plot our results at $r<=0.05$ pc using solid lines. For $r>=0.05$ pc we use dashed lines to indicate regions where scattering may increase the ionization rates and the curve represents a lower limit. 

\subsection{Number of Particles}
\label{appendix_subsection:nparticles}

Statistical counting errors are inherent in Monte Carlo methods. In our fiducial runs with $\sigma^2=0.0$, this effect is negligible because there is only random sampling for the initial particle directions, and the rest of the particles' motion is dictated by the bulk magnetic field. For $\sigma^2>0.0$, sampling the particle direction and step size creates a source of statistical noise that may effect our results for the particle flux and ionization rates around the star.

For each of our simulations we use $N_{PE}=10^{3}$ particles starting in each energy bin, which gives a total particle count of $N=5\times 10^{4}$ particles. Here, we demonstrate that this is enough particles to achieve convergence in our results for the ionization rate in different regions of the core. We analyze our results for $t=0.3$ Myr using $\sigma^2=0$ and $L_c=10^{-5}$ pc, our highest turbulence run, using different values of $N_{PE}$. We define the relative error in the ionization rate as $\delta\zeta_{2000} (N_{PE})=|\zeta(N_{PE})-\zeta_{2000}|/\zeta_{2000}$ where $\zeta_{2000}$ is the ionization rate for $N_{PE}=2\times 10^{3}$ particles. We calculate this quantity for the average ionization rate in the outflow and the maximum ionization rates computed over the whole grid, as these quantities have fewer particle counts and consequently are more sensitive to changes in $N_{PE}$ than the radial average ionization rate.

The right panel of \ref{fig:crir_plots_uncertainties} shows 
our results. As can be seen, the average ionization rates computed in the outflow region are not very sensitive to $N_{PE}$. All values of $\zeta(N_{PE})$ tested have an uncertainty in the outflow ionization rate of less than 100 $\%$ throughout the core, and $\zeta(10^{3})$ has an uncertainty of less than 50 $\%$ throughout the core. Computing an ionization rate in a single angular bin, such as for the maximum ionization rates shown by the green points, have a higher uncertainty. This is in part a function of how many solid angle bins we use, and the uncertainty would be lower if we use fewer bins. For our grid, $\delta\zeta(10^{3}) \approx 1$ at the edge of the core, corresponding to a factor of two discrepancy. All the same, given the magnitude of the uncertainties in our setup, this is a minor effect.  

\section{Impact of the Correlation Length}
\label{appendix:turbulence_additional}

In this section, we explore the impact of the value of the correlation length on our results. As discussed in Section \ref{subsubsection:crparameters}, the value of the correlation length of turbulence in molecular clouds is uncertain, but the limitations of our sampling require that we test parameters that give path lengths roughly the size of a grid cell or smaller. For the simulations presented in Figure \ref{fig:flux_crir_turbulence_13_5} we test $L_c=10^{-5}$ pc, which fits this criteria for our particle energy range. We also run our simulations for $L_c=10^{-4}$ pc, which produces path lengths that approach the cell size but is still small enough to give accurate results. 

\begin{figure*}
\centering
\includegraphics[width=0.98\linewidth]{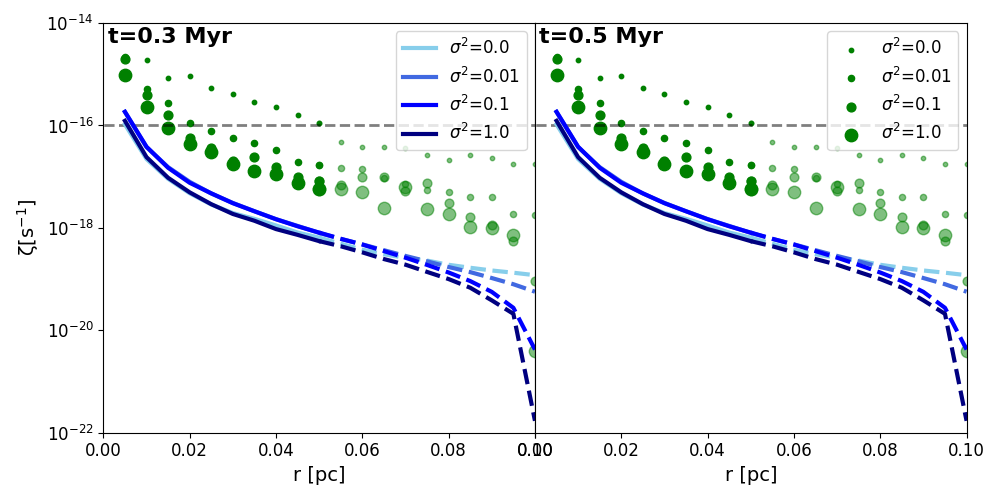}
\caption{Average ionization rate as a function of radius for $t=0.3$ Myr and $t=0.5$ Myr using $\rm L_c=10^{-4}$ pc. Turbulent power is indicated by the shade of blue. Maximum values of any angular bin at a given radius are shown by the green points, with larger dots symbolizing more turbulence. Dashed lines and translucent points at $r>0.05$ pc indicate the region where our results may be uncertain due to our simulation boundary conditions. The gray line indicates $\zeta=10^{-16} \rm s^{-1}$, which is the measured ionization rate for the Milky Way.}
\label{fig:flux_crir_turbulence_14_5}
\end{figure*}

Figure \ref{fig:flux_crir_turbulence_14_5} shows the average ionization rates at $t=0.3$ Myr and $t=0.5$ Myr for different values of the turbulent power $\sigma^2$ using $L_c=10^{-4}$ pc. The ionization rates have the same features as most curves in Figure \ref{fig:flux_crir_turbulence_13_5} show- increasing $\sigma^2$ decreases the maximum ionization rate computed at each radius, but has little effect on the radially averaged ionization rates compared to $\sigma^2=0$. This makes intuitive sense when considering Equation \ref{eq:lambda_sc_2}; a larger correlation length decreases the effect of the turbulence, meaning that we would expect even less change for these simulation runs from the $\sigma^2=0$ case than is shown in Figure \ref{fig:flux_crir_turbulence_13_5}. 

If we tested our simulations with a small enough value of the correlation length, our results would show similar features to the $\sigma^2=1$ curve in the left panel of Figure \ref{fig:flux_crir_turbulence_13_5} because of the increase in turbulence. However, because of the computational expense of such a calculation, we leave this for another study.

\section{Analytical Solution Tests}
\label{appendix:tests}

Many CR propagation codes treat the propagation of CR particles as being either completely free streaming or completely turbulent. To test these limits, we modify the way the particles chose the direction of their next step. For the free streaming tests, all direction changes are turned off, so that the particles propagate away from the star following their initial velocity direction. For the diffusion tests, we set the particles to always sample their direction randomly, rather than sometimes following the direction of the bulk magnetic field. To set the value of $\lambda_{\rm sc}$, we test $\sigma^2=0.01$, $\sigma^2=0.1$ and $\sigma^2=1.0$ using $L_c=10^{-5}$ pc and $\boldsymbol{B_0}(x)=100 \mu $G in every cell. Even though we set a uniform value of $\boldsymbol{B_0}(x)$ in every cell in order to obtain a nonzero value of $\lambda_{\rm sc}$, the direction of it is irrelevant for these tests and thus its implementation differs from our normal grid setup.

We test each case using $N_{PE}=10^3$ particles. The solutions for the attenuation are simple and can be derived analytically if particle energy losses are turned off. We test this scenario first. To benchmark our calculations with a model that includes energy losses, we use the model developed by \cite{silsbee_2019_model}. 

\subsection{Solutions Without Energy Losses}
\label{appendix_subsection:tests_no_losses}

\begin{figure*}
\centering
\includegraphics[width=0.98\linewidth]{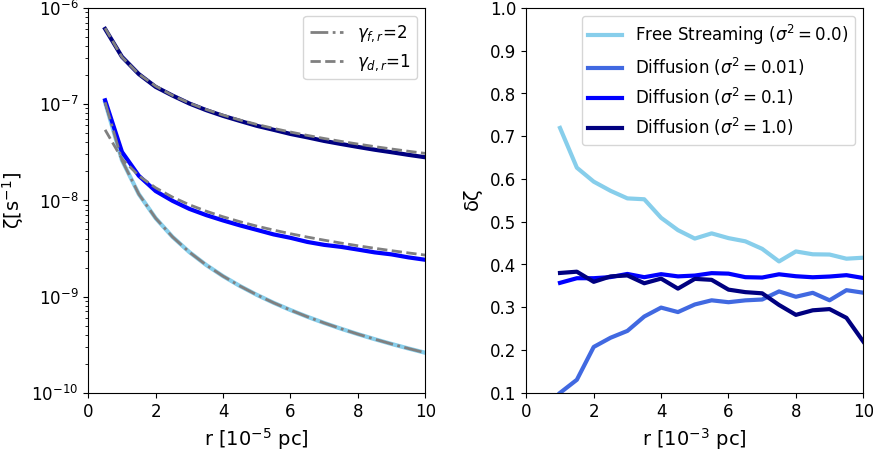}
\caption{Left: Ionization rate for free streaming and diffusion with no energy losses. Dashed lines show comparisons to $\gamma_{f,r}=2$ and $\gamma_{d,r}=1$, which are the analytic solutions for CR attenuation in these limits. Right: Relative error in the ionization rate, $\delta \zeta = |\zeta_{\rm N} -\zeta_{\rm A}|/\zeta_{\rm A}$, for free streaming and diffusion with energy losses, where $\zeta_{\rm A}$ is to the analytic model developed by \cite{silsbee_2019_model}. }
\label{fig:crir_plots_appendix}
\end{figure*}

The simplest tests we perform are for free streaming and diffusive transport where we do not allow the particles to lose energy. As is described in Section \ref{subsection:geometry}, in these limits the attenuation of the CR flux with radius can be derived analytically and follows a power law, $j(E,r)\propto j_{\rm in}(E) r^{-\gamma}$. The values of $\gamma$ are $\gamma_{f,r}=2$ and $\gamma_{d,r}=1$ for free streaming and diffusion respectively. Since the flux is uniformly attenuated at all energies, the ionization rate is also attenuated radially by $r^{-2}$ and $r^{-1}$ for free streaming and diffusion. For these tests we compute the particle trajectories only out to $r=10^{-4}$ pc, because the diffusion tests are computationally expensive with a small value of $\lambda_{\rm sc}$. For each case, we compute the radial ionization rate following Equation \ref{eq:radial_crir}. The left panel of Figure \ref{fig:crir_plots_appendix} shows that our code correctly reproduces the analytic solution in these limits.

\subsection{Solutions With Energy Losses}
\label{appendix_subsection:tests_losses}

We also compare the results of our code in the free streaming and diffusive limits to the model presented in \cite{silsbee_2019_model}, modified to account for the spherical geometry of this test problem. For these tests, we compare our results to exact solutions for the free streaming and diffusive ionization rates. The models take as inputs a power-law exponent of the initial CR spectrum, $a$, and a power-law exponent of the energy loss function, $b$. As discussed in Section \ref{subsection:initial}, the power-law exponent for our initial CR spectrum is $a=1.9$. We use the same loss function as \cite{silsbee_2019_model}, which follows $b=0.82$ in the energy range that is important for ionization ($\approx$ 0.1 MeV- 1 GeV) \citep{padovani_2018_disks}.

The solutions for the attenuation of the ionization rate with column density are functions of the column density $N$. In order to compare to the analytical solutions, we adopt a uniform density of $n(H_2)=10^{7} \rm cm^{-3}$. In this case, $N(r)=n(H_2) r$. The solution for the ionization rate for free streaming propagation is given by
\begin{equation}
    \zeta_{\rm fs}(r)= \frac{1}{4 \pi r^2} \frac{(1+2b)}{(a+2b)}\frac{j_0 L_0 E_0 I_f}{\epsilon}\lp\frac{n(H_2)r}{N_{0f}}\rp^{-\lp\frac{a+b-1}{1+b}\rp},
\end{equation}
where $\epsilon=37$ eV, $I_f \approx 1.07/(a+b-1)+0.42$, and $N_{0f}=\frac{E_0}{L_0(1+d)}$. We choose a reference value of $E_0=1$ MeV, and $j_0$ and $L_0$ are the values of the initial CR spectrum and loss function at $E_0$. The solution for the ionization rate for diffusive propagation is also a function of the power of the turbulence $\sigma^2$. It is given by
\begin{equation}
    \zeta_{\rm diff}(\sigma^2, r)=\frac{1}{\epsilon n(H_2)}\int_{E=0}^{E_{\rm max}}\int_{E'=E}^{E_{\rm max}} \lp\frac{1}{(4\pi K (E'^{\alpha}-E^{\alpha}))^{3/2}}\rp j_0\lp\frac{E'}{E_0}\rp^{-a}\t{exp}\lp\frac{-r^2}{4K(E'^{\alpha}-E^{\alpha})}\rp dE dE', 
\end{equation}
where $\alpha=7/6+b$ for a Kolmogorov spectrum, and 
\begin{equation}
    K=\frac{D_0E_0^{(1-\alpha)}}{\alpha n(H_2) v_0 L_0},
\end{equation}
where $D_0=\lambda_{\rm sc}(E_0) v_0/3$ is the diffusion coefficient at $E_0$.

For these tests, we propagate the particles out to $R_{\rm core}=0.01$ pc and compute the radial ionization rate.  The right panel of Figure \ref{fig:crir_plots_appendix} shows the relative error in the ionization rate as defined $\delta \zeta=|\zeta_{\rm N}-\zeta_{\rm A}|/\zeta_{\rm A}$, where $\zeta_{\rm N}$ and $\zeta_{\rm A}$ are the numerical solution and the analytic solution, respectively. All values of $\zeta_{\rm N}$ have a relative error of less than 75\% throughout the core for all values of $\sigma^2$. Given the orders of magnitude involved in our calculations throughout this study, we consider our results in good agreement with the analytic solution.

\acknowledgements

This work was supported by the US Department of Energy through the Los Alamos National Laboratory. Los Alamos National Laboratory is operated by Triad National Security, LLC, for the National Nuclear Security Administration of U.S.\ Department of Energy (Contract No.\ 89233218CNA000001) This research was supported in part by the National Science Foundation under Grant No. NSF PHY-1748958 and by NASA ATP grant 80NSSC20K0507. This material is based upon work supported by the U.S. Department of Energy, Office of Science, Office of Advanced Scientific Computing Research, Department of Energy Computational Science Graduate Fellowship under Award Number DE-SC0021110. 
We thank an anonymous referee for helpful comments that improved the manuscript.

\bibliography{master}

\end{document}